\documentclass[aps,prb,reprint,showpacs,twocolumn,superscriptaddress]{revtex4-1}
\usepackage[utf8]{inputenc}
\usepackage[american,british]{babel}
\usepackage[T1]{fontenc}
\usepackage[pdftex]{graphicx}  
\usepackage{graphicx, xcolor}
\usepackage{dcolumn}
\usepackage{bm}
\usepackage{amsmath,amsthm,amssymb}
\usepackage{hyperref}
\usepackage{xcolor}
\hypersetup{colorlinks,bookmarksopen,bookmarksnumbered,
citecolor=magenta,
linkcolor=magenta,
pdfstartview=false,
urlcolor=magenta}
\usepackage{verbatim}
\usepackage{algorithm}
\usepackage{ulem}
\usepackage[noend]{algpseudocode}

\definecolor{va}{rgb}{0.0, 0.5, 0.0}

\newcommand{\s}{\sigma}

\begin{document}
\title{Entanglement equipartition in critical random spin chains}

\author{Xhek Turkeshi}
\affiliation{SISSA and INFN, via Bonomea 265, 34136 Trieste, Italy}
\affiliation{The Abdus Salam International Centre for Theoretical Physics, strada Costiera 11, 34151 Trieste, Italy}
\author{Paola Ruggiero}
\affiliation{Department of Quantum Matter Physics, University of Geneva, 24 quai Ernest-Ansermet, 1211 Geneva, Switzerland}
\author{Vincenzo Alba}
\affiliation{Institute  for  Theoretical  Physics,  Universiteit van Amsterdam, Science  Park  904,  Postbus  94485,  1098  XH  Amsterdam,  The  Netherlands}
\author{Pasquale Calabrese}
\affiliation{SISSA and INFN, via Bonomea 265, 34136 Trieste, Italy}
\affiliation{The Abdus Salam International Centre for Theoretical Physics, strada Costiera 11, 34151 Trieste, Italy}

\date{\today}

\begin{abstract}
The reduced density matrix of many-body systems possessing an additive conserved quantity can be decomposed in orthogonal sectors which can be independently analyzed. Recently, these have been proven to equally contribute to entanglement entropy for one dimensional conformal and integrable systems.
In this paper, we extend this equipartition theorem to the disordered critical systems by studying the random singlet phase. 
We analytically compute the disorder averaged symmetry resolved R\'enyi entropies and show the leading orders are independent of the symmetry sector. 
Our findings are cross-checked with simulations within the numerical strong disorder renormalization group.
We also identify the first subleading term breaking equipartition which is of the form $s^2/\ln\ell$ where $s$ is the magnetization of a subsystem of length $\ell$.

\end{abstract}

\maketitle
\section{Introduction} 

Entanglement plays a fundamental role in characterizing quantum phases of matter ~\cite{Amico2007,Calabrese2009R,Laflorencie2015,Eisert2010}. 
For isolated many-body systems at zero temperature, key results have been derived for the entanglement entropy~\cite{Vidal2003,Latorre2004,Calabrese2004,Calabrese2009} and the entanglement spectrum~\cite{Li2008,Lefevre2008,Pollmann2010}. 
For example, the entanglement entropy of gapped one dimensional systems with local interactions follows an area law, 
whereas gapless systems present a logarithmic scaling in subsystem size with a universal prefactor related to the central charge of the underlying 
conformal field theory (CFT).
Entanglement properties have been intensively investigated also in quantum systems with quenched disorder. 
When the low-energy physics of these models is captured by an infinite disorder fixed point, these systems display a 
logarithmic  entanglement entropy scaling which resembles that of a CFT~\cite{Refael2004,Dechiara2006,Laflorencie2005,Refael2007,Hoyos2007,Bonesteel2007,Refael2009,Binosi2007,Getelina2016}, although 
this analogy breaks down in many respects, such as a different scaling of entanglement in other 
circumstances \cite{Igloi2008,Fagotti2011,ruggiero-randneg,Ramirez2014,trc-19} and the
absence of a c-theorem~\cite{Raul2006,Fidkowski2008}.

By partitioning a system into two parts ${A\cup B}$, the bipartite entanglement of a pure state ${|\Psi\rangle}$ is fully encoded in its reduced density matrix ${\rho_A = \text{tr}_B \left(|\Psi\rangle\langle \Psi |\right)}$. 
The spectrum of ${\rho_A}$, known as entanglement spectrum,  can be accessed by studying the scaling of the R\'enyi entropies~\cite{Lefevre2008,act-17}:
\begin{equation}
	\label{eq:2.1.a.v1}
	S_m(\rho_A) = \frac{1}{1-m}\ln \text{tr}_A \left(\rho_A^m\right),
\end{equation}
that for $n\to1$ provide the renowned von Neumann (entanglement) entropy
\begin{equation}	
	\label{eq:2.1.b.v1}
	S(\rho_A) \equiv \lim_{m\to 1} S_m(\rho_A) = -\text{tr}_A \left(\rho_A\log\rho_A\right).
\end{equation}
For one-dimensional quantum systems in the scaling limit, the above can be computed using the replica trick~\cite{Calabrese2004,Calabrese2009R}. 
In {1+1d} CFT (with central charge $c$), explicit results can be obtained in many different situations.
When $A$ is a finite interval of length $\ell$ embedded in an infinite line, one has  the well known formula~\cite{Vidal2003,Latorre2004,Calabrese2004,Calabrese2009}
\begin{align}
	\label{eq:2.2.a.v1}
	S^\textup{CFT}_m(\rho_A) &= \frac{c}{6}\frac{m+1}{m} \ln \ell + \mathit{O}(\ell^0),
\end{align}
The subleading terms are in general non-universal. 
Using conformal transformations, Eq.~\eqref{eq:2.2.a.v1} can be generalized to  finite systems\cite{Calabrese2004}, finite temperature\cite{Calabrese2004}, 
and quench dynamics as well\cite{cc-05}.


Remarkably, the recent technological breakthrough in cold atoms and ion traps lead to high accuracy experiments that directly measure entanglement in these many-body systems~\cite{Islam2015,Kaufman2016,Elben2018,Brydges2019,Lukin2019}. 
Importantly, for a system with additive conservation laws, it is possible to probe different contributions to the entanglement (namely number and configurational entanglement, see Sec.~\ref{sec:symres} for precise definitions) directly related to the entanglement within different symmetry sectors~\cite{Lukin2019}. 
Such symmetry resolution is natural in computational methods like exact diagonalization and tensor network, and has been discussed in earlier papers~\cite{Lauchli2013,Laflorencie2014}. 
In particular, the authors of Ref.~\onlinecite{Laflorencie2014} suggested, using a quantum-thermal correspondence argument, that the entanglement entropy in 
Luttinger liquids is the same for all symmetry sectors. 
Lately, this conjecture has been dubbed \textit{entanglement equipartition} and it has has been proven for conformal\cite{Xavier2018} 
and integrable systems~\cite{Murciano2019}. 

Although there has already been a large interest in the entanglement of quantum systems with internal symmetries for clean systems \cite{Lauchli2013,Laflorencie2014, Goldstein2018,Goldstein2018B,Xavier2018,Feld2019,Bonsignori2019,Murciano2019,Calabrese2020,Bonsignori2019,Murciano2019,Fraenkel2019, crc-20,neg2,clss-19,cms-13,d-16,matsuura,SREE, sara2D},
disordered models lack completely an analytical understanding of the symmetry resolved entanglement spectroscopy (with the notable exception of the 
non-equilibrium experiment in Ref. \onlinecite{Lukin2019}). 
The main question is whether the equipartition of entanglement, shown in a variety of clean models, is robust against the addition of disorder. In this paper we address this issue, by presenting the analytical results for the R\'enyi entanglement entropy in the random singlet phase (RSP).
This class of states characterizes, for instance, the infrared physics of the disordered Heisenberg spin-1/2 chain, and is amenable to exact computations in the thermodynamic limit within the framework of strong disorder renormalization group (SDRG)~\cite{Ma1979,Ma1980,Fisher1994,Igloi05,Monthus2018,Doty1992}. We find that in analogy to clean critical systems, entanglement equipartition holds also for the random singlet phase. Our findings are supported by numerical renormalization group simulations.

In order to maintain this paper self-contained, we first review the symmetry resolved entanglement entropies in Sec.~\ref{sec:symres}. In Sec.~\ref{sec:rsphase} we introduce the model under study, the SDRG method, and the main properties of the RSP.
In Sec.~\ref{sec:ent-rsp} we summarize known results for the scaling of the entanglement in the RSP.
The novel results are presented in Sec.~\ref{sec:srrsp} where we define and study different quantities providing information about the entanglement content of the different symmetry sectors for the RSP.
In Sec. \ref{sec:num} we carefully test our analytic predictions against a numerical implementation of SDRG.
The final section is left for discussion and conclusions, while technical details are contained in one Appendix.

\section{Symmetry-resolved entanglement} 
\label{sec:symres}

Consider a system possessing a global {\it additive} conserved charge $Q$. 
For instance, this symmetry could be abelian such as the total magnetization in spin systems.  
The reduced density matrix of a subsystem can be decomposed into a direct sum of orthogonal sectors. 
To be specific, let us consider a bipartition of the system as  ${A\cup B}$ and a state $\rho$ in a given representation of $Q$. 
The additivity of $Q$ implies that ${Q=Q_A\otimes \mathbf{1}_B +\mathbf{1}_A\otimes Q_B}$ and can be used to show that
\begin{equation}
	\label{eq:2.3.v1}
	 [\rho_A,Q_A]=0.
\end{equation}
Thus the reduced density matrix is block diagonal in the quantum numbers of $Q_A$. 
Denoting with $\Pi_q$ the projector into the subspace relative to the eigenvalue $q$, we have
\begin{equation}
	\label{eq:2.4.v1}
	\rho_A =\oplus_{q}\left( \Pi_q \rho_A \Pi_q \right)=  \oplus_{q} p_A(q) \rho_A(q).
\end{equation}
In the last equality we  factorized the term ${p_A(q) = \text{tr}_A (\Pi_q \rho_A)}$, and defined
\begin{equation}
	\rho_A(q) \equiv \frac{\Pi_q \rho_A \Pi_q }{p_A(q)},\quad \text{tr}(\rho_A(q)) = 1.
\end{equation}
Here $p_A(q)$ is the probability for the subsystem to be in a specific symmetry sector. 
In fact, only the global state possesses a definite charge, 
while the subsystem fluctuates between the $Q_A$-sectors due to quantum effects.

The total von Neumann entanglement entropy of the system 
naturally splits in two parts \cite{Lukin2019,nc-10}
\begin{equation}
\label{eq:2.6.a.v1}
	S(\rho_A)=S^\mathrm{Q} +  S^\mathrm{conf},  
\end{equation}
with
\begin{align}
\label{eq:2.6.v1}
	&S^\mathrm{Q}  = -\sum_q p_A(q)\log p_A(q),\\
	&S^\mathrm{conf}  = \sum_q p_A(q)S(q),
\end{align}
Here ${S(q)\equiv-\mathrm{Tr}\rho_A(q)\ln\rho_A(q)}$ defines the symmetry-resolved entanglement entropy, meaning the contribution to the entanglement entropy of the $q$-sector. 
$S^\mathrm{Q}$ is known as the number (or fluctuation) entropy, since it is related to the 
number of excitations carrying a quantum of symmetry charge, which fluctuates in a subsystem. 
Despite its classical Shannon form, it originates 
from tunneling effects~\cite{Lukin2019}. 
We mention here that the link between entanglement and subsystem's fluctuations (for instance, of spin or particle number in lattice models with a $U(1)$ current) has been widely studied \cite{song2012, song2010, rachel2012,klick2009, song2011,petrescu2014,wv-03,SREE2d,delmaestro2,kusf-20,kusf-20b,cmv-12,si-13,clm-15}.
$S^\mathrm{conf}$ is named configurational entropy, as it 
depends on the many-body coherence pattern of the subsystem 
configurations in a given symmetry sector.  

Similarly, one can define the symmetry-resolved R\'enyi entropies, $S_m(q)$. 
First, we introduce the symmetry-resolved moments:
\begin{equation}
\label{eq:2.7.v1}
	Z_m(q) \equiv 
	{p^m(q)}\mathrm{Tr}(\rho^m_A(q)).
\end{equation}
Note that ${Z_1(q)=p_A(q)}$. Then, we have 
\begin{equation}
\label{eq:2.8.v1}
	S_m(q) = \frac{1}{1-m} \log \left(\frac{Z_m(q)}{Z_1^m(q)}\right).
\end{equation}
They are related to $S(q)$ by the usual limit $m\to 1$.
The symmetry resolved entanglement entropies 
$S_m(q)$ are the main object of study in 
this paper. 

Computing $Z_m(q)$ is in general a non-trivial task. 
A fundamental observation for its derivation is that $Z_m(q)$ is the Fourier transform 
of the charged moment $\mathcal{Z}_m(\alpha) = \text{tr}_A \left(\rho_A^m e^{i Q_A \alpha} \right)$~\cite{Goldstein2018}, i.e.,
\begin{equation}
\label{eq:2.9.v1}
\mathcal{Z}_m(\alpha) = \sum_q e^{i q \alpha}Z_m(q), 
\quad Z_m(q) = \int_{-\pi}^{\pi}\frac{d\alpha}{2\pi} \mathcal{Z}_m(\alpha).
\end{equation}
In some setting, the calculation of ${\mathcal Z}_m(\alpha)$ can be easily performed and then, by Fourier 
transform, symmetry-resolved R\'enyi entropies are obtained. 
This is the case for example for 1+1d CFTs, where the charged moments are easily expressed in path integral language~\cite{Goldstein2018}.
For Luttinger liquids, a particular class of CFTs with central charge ${c=1}$ and characterized by a parameter $K$
(Luttinger parameter), one finds
\begin{equation} \label{eq:Zm_CFT}
Z_m (q) \simeq \ell^{-\frac{1}{6}\left( m-\frac{1}{m}\right)} \sqrt{\frac{n\pi}{2K \ln \ell}} e^{- \frac{n \pi^2 q^2}{2K \ln \ell}},
\end{equation}
leading to
\begin{equation}
	\label{eq:Sm_CFT}
	S_m(q) = S_m^\textup{CFT} - \frac{1}{2}\ln\left( K \ln\ell \right) + \mathit{O}(\ell^0).
\end{equation}
Here $S_m^\textup{CFT}$ is given by~\eqref{eq:2.2.a.v1}. Importantly, Eq.~\eqref{eq:Sm_CFT} shows that
the entanglement entropies of the different symmetry 
sectors are the same at leading orders in the subsystem size 
$\ell$, i.e., Luttinger liquids exhibit entanglement  
equipartition. Corrections to this scaling are in general 
non-universal and model 
dependent~\cite{Bonsignori2019,Murciano2019,Fraenkel2019}. 

\section{Disordered Heisenberg chain and  Random Singlet Phase} 
\label{sec:rsphase}

Here we are interested in the entanglement properties of disordered systems with ground states in the random singlet phase (RSP). 
In the following we introduce the prototypical disordered Heisenberg chain (see~\ref{sec:disH}). 
Its ground-state properties can be addressed by using the Strong-Disorder Renormalisation 
Group (SDRG) method, which we briefly introduce in section~\ref{sec:sdrg}. 

\subsection{Antiferromagnetic Heisenberg spin chain}
\label{sec:disH}

The spin-$1/2$ antiferromagnetic Heisenberg chain 
is defined by the Hamiltonian 
\begin{equation}
\label{eq:3.1.v1}
H_L = \sum_{i=1}^{L} J_i (S_i^x S_{i+1}^x +S_i^y S_{i+1}^y + \Delta S_i^z S_{i+1}^z),
\end{equation}
where $\Delta$ is the anisotropy parameter, and $S_i^{x,y,z}$ 
are spin-$1/2$ operators. 
We restrict ourselves to 
$\Delta=1$ (isotropic Heisenberg chain). Here $J_i$ are positive 
random couplings distributed according to a given distribution $P(J)$. 
In the absence of disorder, i.e., ${P(J)\sim \delta(J-J_0)}$ for some fixed value $J_0$, the ground state of the model is in 
a Luttinger liquid phase at any $-1<\Delta\le 1$. Thus, the scaling 
of the ground-state entanglement entropy is described by the CFT 
formula~\eqref{eq:2.2.a.v1} with $c=1$. In the presence of random 
antiferromagnetic couplings $J_i$, the ground state of the system is 
described by an Infinite-Randomness Fixed Point (IRFP),
irrespective of the initial distribution $P(J)$ and, therefore, of the initial disorder strength~\cite{Fisher1994}. 
More generally, all the long-wavelength properties of the disordered Heisenberg chain are expected to be universal. 
The ground state of~\eqref{eq:3.1.v1} is in the random singlet phase (RSP), which is  the simplest example of IRFP. The structure of the RSP can be understood by using the SDRG method. 

\subsection{Strong-Disorder RG and random singlet phase } 
\label{sec:sdrg}

The SDRG is a real-space renormalization group, particularly suited for inhomogeneous (and therefore for disordered) systems.
We now illustrate the decimation procedure which allows us to obtain the low-energy description of our model. 
We start by considering a $4$-sites isotropic 
Heisenberg Hamiltonian (cf.~\eqref{eq:3.1.v1}, with $\Delta=1$). 
We split $H$ as 
\begin{align}
\label{eq:3.2.a.v1}
	H & \equiv H^{(0)}_{2}+ H^{(1)}_{2},\\
\label{eq:3.2.b.v1}
	H^{(0)}_{2} &=  \Omega \vec{S}_2 \vec{S}_3, \\
\label{eq:3.2.c.v1}
	H^{(1)}_{2} &=  J_L \vec{S}_1 \vec{S}_2+ J_R \vec{S}_3 \vec{S}_4.
\end{align}
Here we assume that ${\Omega> J_L, J_R}$ is the strongest coupling. 
Then we can treat the Hamiltonian $H^{(1)}_{2}$ as a perturbation 
to $H^{(0)}_{2}$. The spins in sites $(2,3)$ bond forming a singlet (the local ground state)
\begin{equation}
|s\rangle = \frac{
\left|\uparrow\downarrow
\right\rangle - \left|\downarrow\uparrow
\right\rangle
}{\sqrt{2}}.
\end{equation}
This also provides an effective coupling for spins $(1,4)$.
Indeed, using second order perturbation theory we 
obtain an effective Hamiltonian $H^\mathrm{eff}_2$ 
for spins $(1,4)$ 
\begin{align}
\label{eq:3.3.a.v1}
	H^\textup{eff}_{2} &= \langle s | H^{(0)}_{2}+ H^{(1)}_{2} | s \rangle + \sum_{t} \frac{|\langle t | H^{(1)}_{2} | s \rangle|^2}{E_s-E_t}\\
\label{eq:3.3.b.v1}
	&= E_0 + \tilde{J} \vec{S}_1\vec{S}_4,\qquad \tilde{J} = \eta \frac{J_L J_R}{\Omega},
\end{align}
where the sum is over the triplet states of two spins, $|t\rangle=|\uparrow\uparrow\rangle,
|\downarrow\downarrow\rangle,(|\uparrow\downarrow\rangle+
|\downarrow\uparrow\rangle)/\sqrt{2}$, and ${E_t=1/4\Omega}$, ${E_s=-3/4\Omega}$.  
 $E_0$ is an unimportant energy constant, 
and  ${\eta=1/2}$ for the isotropic Heisenberg chain. Note that $H^\textup{eff}_{2} $ in \eqref{eq:3.3.b.v1} is still Heisenberg-like, so that the previous steps translate in an effective renormalization of the couplings.

\begin{figure}
\includegraphics[width=\columnwidth]{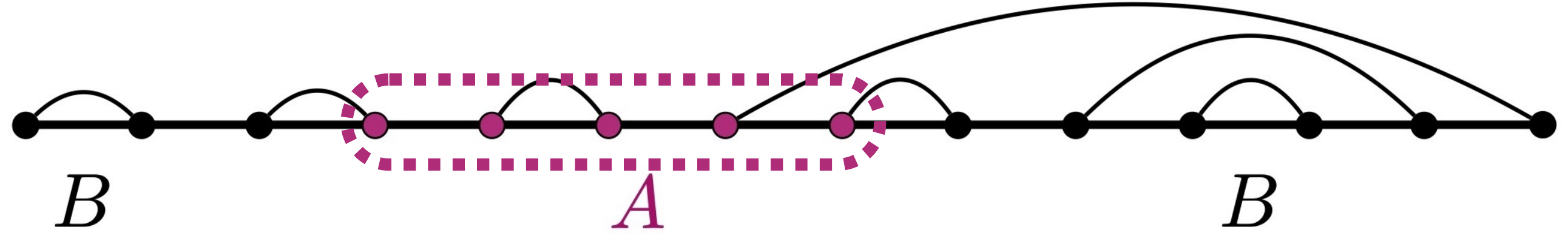}
\caption{Pictorial illustration of the Random Single Phase 
 (RSP). The links connect the spins forming a singlet. The bipartition of the 
 chain as $A\cup B$ is also shown. The entanglement entropy $S$ 
 is proportional to the number of singlets shared between $A$ and $B$. 
\label{fig:rsp_ex}}
\end{figure}

The procedure can then be easily generalized to a many-body hamiltonian with $L$ spins such as Eq.~\eqref{eq:3.1.v1}. 
At each renormalization step, the pair interacting through the strongest coupling ${\Omega=\max\{J_i\}}$ 
forms a singlet which is decimated, and the set of couplings changes according to 
\begin{equation}
	\label{eq:3.4.v1}
	\left(\dots,J_L,\Omega, J_R,\dots\right)_L \rightarrow \left(\dots,\eta \frac{J_L J_R}{\Omega}, \dots\right)_{L-2}.
\end{equation}
This is known as Dasgupta--Ma rule~\cite{Ma1979,Ma1980}.
The RG terminates when all sites are decimated.  The resulting 
state, known as RSP, is a product of singlets ranging arbitrary far in the system and approximates the ground state of the system. Its structure is the same irrespective of 
the chain anisotropy $\Delta$, i.e., chains with different $\Delta$ 
belong to the same universality class. This is illustrated in Fig.~\ref{fig:rsp_ex}. 

\section{Entanglement scaling in Random Singlet Phases} 
\label{sec:ent-rsp}

Here we discuss the entanglement structure of the random singlet phase. 
Given a bipartition ${A\cup B}$  for the chain of even length $L$, 
with $A$ the subsystem of interest (see Fig.~\ref{fig:rsp_ex}), 
the ground state density matrix $\rho_{\mathrm{RSP}}$ is  
obtained as the tensor product of the 
density matrices $\rho_\textup{2s}$ associated with each singlet, resulting in 
\begin{equation}
\label{eq:3.5.v1}
\rho_\textup{RSP}= \bigotimes_{m=1}^{L/2}\rho_\textup{2s}= \bigotimes_{m=1}^{n_{A:A}}\rho_\textup{2s}\bigotimes_{n=1}^{n_{A:B}}\rho_\textup{2s}\bigotimes_{l=1}^{n_{B:B}}\rho_\textup{2s}.
\end{equation}
Here $n_{X:Y}$ is the number of singlets with one end in $X$ and the other in $Y$. 
In~\eqref{eq:3.5.v1} the three different terms correspond to singlets formed by 
spins in $A$ in $B$, or shared between them. 
The singlet density matrix $\rho_\textup{2s}$ reads 
\begin{equation}
\label{eq:3.6.v1}
	\rho_\textup{2s} = \frac{1}{2}
	\left(\left|\uparrow\downarrow
\right\rangle - \left|\downarrow\uparrow
\right\rangle\right)
\left(\left\langle \uparrow\downarrow\right| - \left\langle \downarrow\uparrow\right| \right).
\end{equation}
The trace over $B$ does not affect the $n_{A:A}$ singlets 
within $A$. Instead, after tracing over $B$, each of the spins of 
the $n_{A:B}$ shared singlets is described by the 
mixed-state reduced density matrix $\rho_{\textrm{s}}$
\begin{equation}
\rho_{\textrm{s}}= \frac{1}{2}(\left|\uparrow\right\rangle\langle 
\left\uparrow\right|+\left|\downarrow\right\rangle\langle \left\downarrow\right|), 
\end{equation}
which is diagonal with two equal eigenvalues $1/2$. 
Thus, the reduced density matrix is
\begin{equation}
\label{eq:3.7.v1}
	\rho_A = \bigotimes_{m=1}^{n_{A:A}}\rho_\textup{2s}\bigotimes_{n=1}^{n_{A:B}}\rho_\textup{s}.
\end{equation}
The singlets created within $A$ do not contribute to the entanglement 
between $A$ and $B$, which is obtained by the second term in~\eqref{eq:3.7.v1}. 
For later convenience, let us define $\rho_{\mathrm{in/out}}$ as 
\begin{equation}
\label{eq:inout}
\rho_\mathrm{in/out} = \bigotimes_{n=1}^{n_{A:B}}\rho_\textup{s}.
\end{equation}
The entanglement spectrum, i.e., the eigenvalues of $\rho_A$ is fully 
characterized by the in-out singlets ${n_{A:B}}$, 
which constitute the Bell pairs between the parties $A$ and $B$. 
The contribution of each shared singlet to any R\'enyi entanglement entropy is $\ln 2$, and thus
\begin{equation}
	\label{eq:vn-nab}
	S_m=S=n_{A:B}\ln 2,
\end{equation}
valid for each disorder realization. 

Clearly, in a disordered model, as $n_{A:B}$ fluctuates in different disorder 
realizations, what is meaningful is its average over many realizations. 
This can be characterized through the SDRG approach. 
For the mean value $\langle n_{A:B} \rangle $ of a subsystem consisting of an interval $A$ of length $\ell$ one obtains (we refer to Refs.~\onlinecite{Refael2004,Refael2009} for a detailed derivation)
\begin{equation}
\label{eq:3.8.v1}
	\langle n_{A:B} \rangle  = \frac{1}{3}\ln\ell + \mathit{O}(\ell^0).
\end{equation}
More generally speaking, one can consider the generating function $g(\s)$ of all moments of ${n}_{A:B}$, defined as 
\begin{equation}
\label{eq:3.11.v1}
g(\s)= \sum_{n_{A:B}} P(n_{A:B}) e^{-n_{A:B}\s},
\end{equation}
where $s$ is  real parameter and $P(n_{A:B})$  is the full distribution 
of the shared singlets. 
Specifically, from \eqref{eq:3.11.v1},
\begin{equation}
	\langle n^k_{A:B} \rangle =\left.\frac{d^kg(\s)}{d\s^k}\right|_{\s=0}.
\end{equation}
Note that after replacing the summation 
with an integral, Eq.~\eqref{eq:3.11.v1} is the Laplace 
transform of $P(n_{A:B})$. 

Importantly, in the scaling limit of large $\ell$, $g(\s)$ can be calculated 
within the SDRG framework (by using a renewal-equation approach, 
see Ref.~\onlinecite{Fagotti2011} for details).  
The result reads 
\begin{multline}
	g(\s) = e^{-3/2 \mu} \Bigg[ \cosh\left(\frac{\sqrt{5+4e^{-\s}}}{2}\mu \right)\\ \label{eq:g} 
		+\frac{3}{\sqrt{5+4e^{-\s}}}\sinh\left(\frac{\sqrt{5+4e^{-\s}}}{2}\mu \right) 
	\Bigg], 
\end{multline}
where
\begin{equation}
	\label{eq:mu}
	\mu = 3 \langle n_{A:B} \rangle + \frac{1}{3}.
\end{equation}
Eq.~\eqref{eq:g} holds in the scaling 
limit ${\ell\to\infty}$. Away from the scaling limit, 
corrections due to finite $\ell$ are expected~\cite{Fagotti2011}.

We now discuss the consequences for the entanglement entropies.
In particular, we introduce two different definitions of disorder-averaged 
entropies. In the first, the average over the disorder is 
taken after the logarithm of the moments of the reduced 
density matrix, i.e., we average Eq.~\eqref{eq:2.1.a.v1}. This defines the entropies  $\overline{S}$ 
and $\overline{S}_m$ as 
\begin{align} \label{barSvN}
& \overline{S}\equiv - \langle {\mathrm{tr}_A\rho_A\ln\rho_A} \rangle, \\
\label{barSm}
& \overline{S}_m\equiv \frac{1}{1-m}\langle {\ln\mathrm{tr}_A\rho_A^m} \rangle . 
\end{align}
with ${\langle \bullet \rangle =\sum_{n=0}^{\infty} P(n)\,  \bullet \,}$ being the disorder average.
From~\eqref{eq:3.7.v1}, it is straightforward to see that there is a trivial dependence on the R\'enyi index $m$, i.e, 
\begin{equation} 
\label{eq:barSvNm}
\overline{S}_m=\overline{S}  \quad\forall m.
\end{equation}
Moreover, Eq.~\eqref{eq:vn-nab} depends
only on the average number of shared singlets $n_{A:B}$. Therefore, by using Eq.~\eqref{eq:3.8.v1}, we get
\begin{equation}
\label{eq:3.9.v1}
\overline{S}_m = \langle {n}_{A:B} \rangle  \ln 2 = \frac{\ln2}{3}\ln\ell+ \mathit{O}(\ell^0).
\end{equation}

The disorder averaged version of eq.~\eqref{eq:2.1.a.v1},
is not sufficient to study the full entanglement spectrum~\cite{Fagotti2011}. 
To further investigate the entanglement structure of random singlet phase, it is custom to define
\begin{align}
\label{eq:3.10.v1}
\widetilde{S}_m\equiv\frac{1}{1-m}\ln \langle  {\mathrm{tr}_A\rho_A^m} \rangle,
\end{align}
where the average is taken before the logarithm, i.e., it is the logarithm of the averaged partition function.
Now, $\widetilde{S}_m$ depend on the full distribution of 
in-out singlets $P(n_{A:B})$, which encodes the full entanglement 
content of the random singlet phase. 
Note that, by making use of the following identity
\begin{equation} \label{limlog_der}
\lim_{m\to 1}\frac{1}{1-m}\ln f(m) =-\partial_m f(m) |_{m=1},
\end{equation}
valid for any function $f(m)$ such that ${f(1)=1}$, one can show that, in the limit ${m\to1}$, $\widetilde{S}_m$ and $\overline{S}_m$ coincide, i.e. 
\begin{equation}
\label{eq:mto1}
\lim_{m\to1}\widetilde{S}_m=\lim_{m\to1}\overline{S}_m=\overline S.
\end{equation}
More generally, from the definition of $g(s)$ in Eq.~\eqref{eq:3.11.v1}, 
it is straightforward to obtain the R\'enyi entropies in Eq.~\eqref{eq:3.10.v1} as
\begin{equation}
\label{eq:3.12.v1}
	\widetilde{S}_m = \frac{1}{1-m} \ln g((m-1)\ln 2).
\end{equation}
By using $g(s)$ in Eq.~\eqref{eq:g}, we obtain 
\begin{equation}
\label{eq:3.14.v1}
	\widetilde{S}_m = 
	\frac{\sqrt{5+2^{3-m}}-3}{2(1-m)}\ln\ell + \textit{O}(\ell^0).
\end{equation}
The subleading term is non-universal and disorder dependent. Importantly, from \eqref{eq:3.14.v1},
we recover
\begin{equation}
\label{eq:3.15.v1}
	 \lim_{m\to 1} \widetilde{S}_m = 
	\frac{\ln 2}{3}\ln\ell+\textit{O}(\ell^0),
\end{equation}
which is consistent with~\eqref{eq:mto1}. 

We mention that SDRG methods can be used also to derive predictions for the entanglement scaling in other phases of matter 
more complicated than RSP \cite{Lin2007,Yu2008,Kovaks2009,Kovaks2012,Laguna2016,asr-18,pcp-19,m-20,vrhs-17,vjs-13}.

\section{Symmetry resolved entanglement in the random singlet phase}
\label{sec:srrsp}

In this section we study the symmetry-resolved entanglement in the random singlet phase. 
In particular, we focus again on the disordered Heisenberg chain (cf.~Eq.~\eqref{eq:3.1.v1}), even if the following discussion can be adapted to all other models 
in the RSP possessing an additive symmetry.

Also in the presence of disorder, the Heisenberg chain for arbitrary $\Delta$ is 
$U(1)$ symmetric because of the conservation of the total magnetization 
${{S}_\mathrm{tot}^z=\sum_i S_i^z={S}_A^z\otimes  \mathbf{1}_B+\mathbf{1}_A\otimes {S}_B^z}$. 
Indeed each hamiltonian term in \eqref{eq:3.1.v1} commutes with ${S}_\mathrm{tot}^z$, i.e., 
\begin{equation}
[J_i (S_i^x S_{i+1}^x +S_i^y S_{i+1}^y + \Delta S_i^z S_{i+1}^z), S^z_i+S^z_{i+1}]=0.
\end{equation}
At the isotropic point, the symmetry is enlarged to $SU(2)$, but here we focus on the more ubiquitous $U(1)$ symmetry.

In the following, we introduce the shorthand ${n\equiv n_{A:B}}$, with, as usual, $n_{A:B}$ the 
number of singlets shared between $A$ and $B$. 
In order to find the internal structure of $\rho_A$,  the first trivial observation is that the singlets within $A$ do not contribute to the 
subsystem magnetization. 
Hence, the possible values of such magnetization only depend on shared singlets. 
Each shared singlet can provide either $+1/2$ or $-1/2$ and consequently the $(n+1)$ possible values
which we denote by $s$ are ${s\in\{-n/2, -n/2+1, \dots,n/2\}}$.
Consequently, throughout this and next section, $s$ stands for the possible eigenvalue of the conserved charge within $A$, i.e., the 
quantity denoted by $q$ in Sec. \ref{sec:symres}.

As it should be clear at this point, entanglement properties of the random singlet phase can be extracted by only looking at the reduced density matrix $\rho_\mathrm{in/out}$ (cf.~\eqref{eq:inout}).
This is true in particular for the contributions from different symmetry sectors.
Now, $\rho_\textup{in/out}$ is of size 
${2^n\times 2^n}$ and has a block structure, 
with ${(n+1)}$ blocks with charge ${s\in \{-n/2,\dots,n/2\}}$. 
The dimension of the block corresponding to $s$ is  
\begin{equation}
\label{eq:4.3.v1}
	d_{s} = \binom{n}{n/2+s}.
\end{equation}
The sum rule $\sum_s d_s=2^n$ holds true from Newton binomial theorem. 

The main ingredient to study the symmetry-resolved entanglement 
is the resolved partition function $Z_m(s)$ defined as
\begin{equation}
\label{eq:4.2.v1}
Z_m(s) \equiv \mathrm{Tr}\left( \Pi_s
\rho_A^m\right)= \mathrm{Tr}\left( \Pi_s
\rho^m_\textup{in/out}\right), 
\end{equation}
where $\Pi_s$ here denotes the projection in the sector with magnetization $s$. 
In the singlet basis, all the blocks of $\rho_\mathrm{in/out}$ are diagonal with equal diagonal elements $2^{-n}$. 
Consequently,  a simple computation gives the symmetry-resolved moments \eqref{eq:4.2.v1} as 
\begin{equation}
\label{eq:4.4.a.v1}
	Z_m(s) = d_{s} 2^{-m n} = \binom{n}{n/2+s} 2^{-m n }. 
\end{equation}
We recall that, in terms of $Z_m(s)$, the symmetry-resolved R\'enyi entropies are 
\begin{equation}
\label{sm_symm_def}
S_m(s) =\frac{1}{1-m}\ln\left[ \frac{Z_m(s)}{Z^m_1(s)} \right],
\end{equation}
which holds {\it for a given disorder realization}.

Next we want to consider the corresponding disorder average. From Eq.~\eqref{sm_symm_def}, it is clear that we can take a few different averages. 
Specifically, we introduce three different quantities in the following. 
We recall that, in the singlet language, the explicit meaning of the average is ${\langle \bullet \rangle =\sum_{n=0}^{\infty} P(n)\,  \bullet \,}$.
 
The first one is 
\begin{equation}
	\label{eq:4.5.a.v1}
	\overline{S}_m(s) = \frac{1}{1-m} \left\langle {\ln \left[ \frac{Z_m(s)}{Z_1^m(s)} \right] } \right\rangle ,\\[5pt]
\end{equation}
which is a genuine average of the symmetry resolved entropies. 
It is analogous to Eq. \eqref{barSm} for each symmetry sector. 
Although this is the most natural quantity, it is the less interesting one from a theoretical perspective, i.e., 
from the point of view of the information that is encoded into it. 

The second one is 
\begin{equation}
	\label{eq:4.5.b.v1}
	\widetilde{S}_m(s) = \frac{1}{1-m} \ln {\left\langle \frac{Z_m(s)}{Z_1^m(s)}\right\rangle},
\end{equation}	
which represents the logarithm of the average of the $m$-th moment. 
It is the symmetry resolved version of Eq. \eqref{eq:3.10.v1} and it is the most suitable quantity to access the symmetry resolved spectrum. 

Finally we also have
\begin{equation}
	\label{eq:4.5.c.v1}
	\widehat{S}_m(s) = \frac{1}{1-m} \ln \left[ \frac{\langle{Z}_m(s) \rangle}{\langle{Z}_1(s)\rangle^m} \right],
\end{equation}
which is the ratio of the averages of the symmetry resolved partitions.  
$\widehat{S}_m(s)$ has no equivalent in the definitions of total entropies, but it is the quantity naturally related to 
the Fourier transforms of charged entropies, as we shall also see in more details in the following. 
Hence it is the average that is closely related to clean systems.

Our main result, which we are going to show soon, is that all the entropies defined in 
Eqs.~\eqref{eq:4.5.a.v1}, ~\eqref{eq:4.5.b.v1} and \eqref{eq:4.5.c.v1} satisfy  
the same equipartition law for the leading and first subleading orders for large subsystem size $\ell$. 
The violations of entanglement equipartition at higher-order are non-universal. 

Before proceeding, some observations are in order to set up the calculations. 
First, plugging Eq.~\eqref{eq:4.4.a.v1} into Eq.~\eqref{eq:4.5.a.v1}, we obtain that $\overline{S}_m(s)$ 
reads as 
\begin{equation}
\label{eq:barS}
\overline{S}_m(s) = \left\langle \ln \binom{n}{n/2+s}\right\rangle.
\end{equation}
Note that similarly to Eq.~\eqref{eq:3.9.v1},  $\overline{S}_m(s)$ does not 
depend on $m$. For $\widetilde{S}_m(s)$, plugging Eq.~\eqref{eq:4.4.a.v1} into Eq.~\eqref{eq:4.5.b.v1}, one has the similar expression 
\begin{equation}
\label{eq:4.5.b.v2}
\widetilde{S}_m(s) = \frac{1}{1-m} \ln \left\langle \binom{n}{n/2+s}^{(1-m)}\right\rangle.
\end{equation}
Using Eq.~\eqref{limlog_der}, it is straightforward to see that Eq.~\eqref{eq:4.5.b.v2} and~\eqref{eq:barS} coincide in the limit ${m\to1}$, leading to
\begin{equation} \label{eq:Srsp_numerics}
\overline{S}(s)= \widetilde{S} (s)= \left\langle \ln \binom{n}{n/2+s}\right\rangle.
\end{equation}
This result is also in full analogy with the total entropy where the two limits of Eq. \eqref{barSm} and \eqref{eq:3.10.v1} coincide at $m=1$.

Conversely, $\widehat{S}_m(s)$ provides a different limit for $m\to1$, 
Indeed, plugging Eq.~\eqref{eq:4.4.a.v1} into Eq.~\eqref{eq:4.5.c.v1} we have 
\begin{equation}
\widehat{S}_{m} (s)= \frac1{1-m}\ln  \frac{\left\langle  2^{-m n} \binom{n}{n/2+s} \right\rangle }{\left\langle 2^{-n} \binom{n}{n/2+s} \right\rangle^m} ,
\label{Shatav}
\end{equation}
that in the limit ${m\to1}$ becomes
\begin{multline} \label{hatS_vN}
\widehat{S} (s)=  \frac{\left\langle n \, 2^{-n} \binom{n}{n/2+s} \right\rangle}{\left\langle 2^{-n} \binom{n}{n/2+s}\right\rangle} \ln 2 
+\ln \left\langle 2^{-n} \binom{n}{n/2+s}\right\rangle,
\end{multline}
where we used again the identity \eqref{limlog_der}.
Finally, it is important to notice that only the calculation of $\widehat{S}_m(s)$ 
involves explicitly $\langle {Z}_m (s) \rangle$.  


\subsection{Preliminaries}

Before embarking into the specific calculations of the various entropies, we discuss the asymptotic limit in which we are interested and
the simplifications taking place in such a limit.
First of all, we observe that the main ingredient for the computation of symmetry resolved entropies 
are the averaged integer (negative) powers of the size $d_s$ of block at fixed symmetry resolution $s$. 
We introduce the quantity $I_m(s)$ as  
\begin{equation}
\label{eq:4.6.v1}
	I_m(s) \equiv \langle d_s^{1-m}\rangle= \sum_{n=0}^{\infty} P(n) \binom{n}{n/2+ s}^{(1-m)},
\end{equation}
which is directly related to the entropy $\widetilde S_m(s)$ as
\begin{equation}
\widetilde{S}_m(s)= \frac{\ln (I_m(s))}{1-m}.
\end{equation}

The calculation of the average $I_m(s) $ at finite size is a hard task (likely impossible), since it requires a precise knowledge 
of the probability distribution $P(n)$. 
However, we are only interested in the scaling limit with large $\ell$. In this case, the mean number of singlets $\langle n\rangle$ is large (cf. Eq~\eqref{eq:3.8.v1}). 
Thus, for the average in Eq.~\eqref{eq:4.6.v1}, we can focus on the large $n$ limit, using 
the Stirling approximation to expand the Newton binomial in Eq.~\eqref{eq:4.6.v1} to obtain:
\begin{equation}
\label{eq:expansion}
	\binom{n}{n/2+s}^{1-m} = 2^{n(1-m)}\left(\frac{\pi n}{2}\right)^{\frac{m-1}{2}} 
	\sum_{k=0}^\infty Q_k(n)\left(\frac{s}{n}\right)^2,
\end{equation}
where $Q_k(n)$ are algebraic functions in $n$. For instance, one has:
\begin{align}
\label{eq:q0}
	Q_0(n) &= 1 + {O}\left(\frac{1}{n}\right),\\
\label{eq:q1}
	Q_1(n) &= 2(m-1) n + {O}\left(1\right).
\end{align}
We are interested in small values of $s$ since they are those with a significant contribution to the total entropy (the probability $p(s)$ is expected to decay very quickly with $s$, as we self-consistently show). Then, in the limit of ${n\gg s}$, the only relevant term is the one with $k=0$ in Eq. \eqref{eq:expansion}, i.e.,
\begin{equation}
\label{eq:expansion0}
\binom{n}{n/2+s_q}^{1-m}\simeq
\left({\frac{\pi n}{2 }}\right)^{\frac{m-1}{2}} 2^{n(1-m)} .
\end{equation}

Recalling that $g(\s)\equiv \langle e^{-n \s}\rangle$ (cf. Eq. \eqref{eq:3.11.v1}), the average over the disorder of Eq. \eqref{eq:expansion0}
for $m$ odd is straightforwardly related to the derivative of $g(\s)$. %
Consequently we have 
\begin{multline}
\label{dertrick}
I_{m}(s)\simeq \langle 2^{n(1-m)} (\pi n/2)^{\frac{m-1}{2}} \rangle=\\
= \Big(\frac{\pi}2\Big)^{\frac{m-1}{2}} \frac{\partial^a g(\s)}{\partial (-\s)^a}  \bigg|_{\s=(m-1)\ln2, a=\frac{m-1}2} .
\end{multline}
One can easily perform explicitly the $a$-th derivative (at leading order), obtaining after simple algebra 
\begin{multline}
\label{eq:4.10.v1}
I_m(s)\simeq I^{(0)}_m\equiv
\frac{1}{2}e^{-\frac{1}{2}(3-\sqrt{5+2^{3-m}})\mu}\\
\times 
\Big(1+\frac{3}{\sqrt{5+2^{3-m}}}\Big) \Big(\frac{\pi 2^{-m}\mu}{\sqrt{5+2^{3-m}}}\Big)^{\frac{m-1}{2}}, 
\end{multline}
where ${\mu=\ln\ell}+\dots$ is the same as in Eq. \eqref{eq:mu}. 
At this point, we have an analytic expression for odd $m$. 
It is very reasonable to assume that the same expression indeed provides the correct result for even $m$.
An explicit calculation valid for arbitrary real $m$ can be performed by exploiting the Laplace transform of Eq. \eqref{eq:4.6.v1} for large $n$. 
The calculation is very cumbersome, although it employs only standard techniques of complex integration. 
To maintain a clear exposition of our results, we report the details in Appendix~\ref{app:comp} and just state here that 
such a complex calculation reproduces Eq. \eqref{eq:4.10.v1} at the leading order (but suggests that some deviations are present at 
subleading ones).

Let us quickly discuss what Eq. \eqref{eq:expansion} suggests for the correction to the leading behavior in  Eq. \eqref{eq:4.10.v1}.
The first corrections comes from the term with $k=1$ that for large $n$ multiplies the leading factor by a term $\propto s^2/n$. 
This implies that such a correction term is proportional to $\langle 2^{n(1-m)} n^{\frac{m-1}{2}-1} \rangle$. 
Once again, using the derivative trick in Eq. \eqref{dertrick}, we quickly obtain that this correction is $\propto I^{0}_m/ \mu$. 
Recalling that $\mu=\ln \ell+O(\ell^0)$, we conclude that
\begin{equation}
I_m(s)=I_m^{(0)}\Big(1+ (m-1) 2^{m} \sqrt{5+2^{3-m}}\frac{s^2}{\ln \ell}+\dots \Big)\,,
\label{corr}
\end{equation}
where we explicitly work out the constant multiplying $s^2/\ln\ell$.  
This analysis suggests that the first term that breaks equipartition in $I_m(s)$ is proportional to $s^2/\ln \ell$. 
This is reminiscent of what observed for clean systems in few different situations \cite{Bonsignori2019,crc-20}. 
However, we must stress that Eq. \eqref{corr} should be taken with a grain of salt. 
Indeed, it assumes the validity of the form \eqref{eq:g} for $g(\s)$ also for the subleading term. 
It is however known that subleading non-universal terms, not encoded in $g(\s)$,
are present; they are model-dependent and more difficult to calculate (see Ref.~\onlinecite{Fagotti2011} for an in-depth discussion).

\subsection{Entanglement equipartition for $\overline{S}_m$}

The first case we consider is the symmetry resolved entropy $\overline{S}_m(s)$ defined in Eq. \eqref{eq:4.5.a.v1} and 
given by the mean value of $\ln d_s$, cf. Eq. \eqref{eq:barS}. 
This logarithm is simply deduced by exploiting 
\begin{equation}
 \left\langle \ln \binom{n}{n/2+s}\right\rangle= -\frac{\partial}{\partial m} I_m(s)\bigg|_{m=1}\,, 
\end{equation}
and using the zeroth order approximation for $I_m(s)$ in Eq. \eqref{eq:4.10.v1}. 
Keeping the subleading terms up to $O(\ell^0)$, we  obtain
\begin{multline}
\label{eq:4.12.v1}
\overline{S}_m(s) = \frac{\ln 2}3 \mu -\frac12 \ln \Big(\frac{\pi}6 \mu\Big)-\frac{\ln 2}9+O(\mu^{-1})=\\
=\overline S -\frac12 \ln \Big(\frac{\pi}6 \mu\Big) +O(\mu^{-1})=
\overline S - \frac{1}{2}\ln\ln\ell+\dots.
\end{multline}
Equation~\eqref{eq:4.12.v1} is the first main result of this paper: it shows the entanglement equipartition of the random singlet phase for the entropy $\overline{S}_m(s)$. 
The leading contributions to these R\'enyi entropies are the same for all the symmetry sectors.  
The first term is just the total entanglement. We will discuss the origin of $-1/2 \ln\ln \ell$ at the end of the section, because the same term 
will appear in all other entropies we consider.  
The $O(1)$ term is not universal, but we reported it here for some comparisons we will do later on.  

Let us briefly discuss the corrections to this leading behavior. 
Exploiting Eq. \eqref{corr}, i.e., considering only those coming from Eq. \eqref{eq:expansion}, we simply have that the first term breaking equipartition
should behave as $s^2/\ln \ell$. 
More quantitatively, from Eq. \eqref{corr} we have
\begin{equation}
\overline S_m(s)- \overline S_m(0)= -{6} \frac{s^2}{\mu}\,. 
\label{corrbar}
\end{equation}
Accordingly, at least at this order and within these approximations, $\overline S_m(s)$ is a monotonous {\it decreasing} function of $|s|$.

\subsection{Entanglement equipartition of $\widetilde{S}_m$} 
\label{sec:equi1}

Here we show the entanglement equipartition for the R\'enyi entropies $\widetilde{S}_m$  for arbitrary $m$.
The integral $I_m(s)$ is directly related to $\widetilde{S}_m(s)$ as $\widetilde{S}_m(s)= (\ln I_m(s))/(1-m)$. 
Hence, we have 
\begin{multline}
\label{eq:4.11.v1}
\widetilde{S}_m(s) = 
\frac{\sqrt{2^{3-m}+5}-3}{2(1-m)} \mu - \frac{1}{2}\ln \frac{2^{-m} \pi \mu}{\sqrt{2^{3-m}+5}}+\\+\frac1{1-m}\ln  \Big(\frac{3}{2 \sqrt{2^{3-m}+5}}+\frac{1}{2}\Big)+O(\mu^{-1})
=\\= \widetilde{S}_m -\frac{1}{2}\ln \frac{2^{-m} \pi \mu}{\sqrt{2^{3-m}+5}}+O(\mu^{-1})=\\
	=\widetilde{S}_m - \frac{1}{2}\ln\ln\ell + \dots.
\end{multline}
The leading logarithmic term is the same as in the total R\'enyi entropy $\widetilde{S}_m$.
Again, there is an additional universal double-logarithmic term in $\ell$ which is not present in the total entropy. 
Both these terms  are independent of the symmetry sector $s$, i.e., the symmetry-resolved 
R\'enyi entropies $\widetilde{S}_m(s)$ exhibit equipartition.
In both Eqs. \eqref{eq:4.11.v1} and \eqref{eq:4.12.v1}, we have been very careful to write the entire subleading term at order $O(\ell^0)$. 
This has been done to show their relationship with the $O(\ell^0)$ terms in the total entropies from Ref. \onlinecite{Fagotti2011}.
We stress however that they have all been obtained with the assumptions used to derive $g(\s)$ in Ref. \onlinecite{Fagotti2011}.

Even for this entanglement measure, from  Eq. \eqref{corr}, we expect subleading logarithmic corrections to Eq.~\eqref{eq:4.11.v1} to violate equipartition as $s^2/\ln \ell$. 
Specifically, from Eq. \eqref{corr} we have
\begin{equation}
\widetilde S_m(s)- \widetilde S_m(0)= -{2^{m}\sqrt{2^{3-m}+5}} \frac{s^2}{\mu}\,.
\label{corrtil}
\end{equation}
Once again, at least at this order and within these approximations, $\widetilde S_m(s)$ is a monotonous {\it decreasing} function of $|s|$ for all values of $m$.

\subsection{Entanglement equipartition for  $\widehat{S}_m$}
\label{sec:sbar}

In this section we explicitly compute $\widehat{S}_m(s)$ and show that 
entanglement equipartition holds also for it. 
Moreover, we will show that also the double logarithnmic term is the same as for $\widetilde{S}_m(s)$ and 
$\overline{S}_m(s)$. 

The main ingredient to compute $\widehat{S}_m(s)$ is the average 
$\langle Z_m (s) \rangle$.  Here the strategy is to first 
evaluate the disorder average of the charged moments 
$\langle {\mathcal{Z}}_m(\alpha) \rangle$ and then to perform a 
Fourier transform (cfr. Eq.~\eqref{eq:2.9.v1}). 
\textit{En passant} this will give access to the probability $p(s)$ characterizing each sector's population. 

In a given disorder realization with $n$ shared singlets, the charged moment ${\mathcal{Z}_m(\alpha)}$ reads
\begin{multline}
	\mathcal{Z}_m(\alpha)= 
	\sum_{s=0}^{n} 2^{-m n} \binom{n}{q} e^{i \alpha (s-n)/2}\\
	= \left[2^{(1-m)}\cos\left(\frac{\alpha}{2}\right)\right]^{n},\label{eq:4.13.v1}
\end{multline}
and consequently  its disorder average is 
\begin{equation}
\label{eq:4.14.v1}
	\langle {\mathcal{Z}}_m(\alpha) \rangle= g\left(2^{(1-m)}\cos\left(\frac{\alpha}{2}\right)\right),
\end{equation}
where $g(\s)$ is the generating function in \eqref{eq:3.11.v1}.
Incidentally for $m=1$, $\langle {\mathcal{Z}}_1(\alpha) \rangle$ is the full counting statistic generating function of 
this disordered model.

Exploiting the explicit knowledge of $g(\s)$ in Eq. \eqref{eq:g}, the Fourier transform of Eq.~\eqref{eq:4.14.v1} can be computed 
by using the saddle point approximation in the scaling limit ${\ell\gg 1}$ (equivalently $\mu\gg1$, cfr. Eq.~\eqref{eq:mu}), obtaining
\begin{multline}
	\langle {Z}_m(s) \rangle= \left(\frac{1}{2}+\frac{3}{2\sqrt{5+2^{3-m}}}\right) e^{-\frac{3}{2}\mu+\frac{\sqrt{5+2^{3-m}}}{2}\mu} \\
	\times \sqrt{\frac{2^m \sqrt{5+2^{3-m}}}{ \pi \mu }}
	\exp\left[-\frac{2^m \sqrt{5+2^{3-m}}}{\mu}s^2\right].	\label{eq:4.15.v1}
\end{multline}
Interestingly, in this approximation $\langle {Z}_m(s) \rangle$ has the very same structure of conformal result, cf. Eq.~\eqref{eq:Zm_CFT}, i.e.,
it is gaussian with variance $\propto \mu$. 
From Eq.~\eqref{eq:4.15.v1} we can directly read out the probability for the subsystem magnetization to be equal to $s$ as  
\begin{equation}
\label{eq:4.16.v1}
	p(s) =\langle {Z}_1(s) \rangle = \sqrt{\frac{6}{\pi \mu}}\exp\left[{-\frac{6 s^2}{\mu}}\right].
\end{equation}

Finally, we plug the partition function \eqref{eq:4.15.v1} into the definition \eqref{eq:4.5.c.v1},
to get the entropy $\widehat{S}_m(s)$ as
\begin{align}
	\widehat{S}_m(s) &= \widetilde S_m(s) +\frac{m}{2(m-1)} \ln \Big(\frac{2^m}6 \sqrt{2^{3-m}+5}\Big) +O(\mu^{-1})\nonumber \\
	&= \frac{\sqrt{2^{3-m}+5}-3}{2-2m} \ln\ell - \frac{1}{2}\ln \ln\ell + \dots\nonumber\\
	\label{eq:4.17.v1}&= \widetilde{S}_m -\frac{1}{2}\ln\ln\ell+ \dots.
\end{align}
We see that at leading universal orders $\widehat{S}_m (s)=\widetilde{S}_m (s)$ (and the same holds in the limit $m\to1$), 
with a non-universal $\mathit{O}(\ell^0)$ difference in the thermodynamic limit. 

We close this subsection with the highlight of a peculiar phenomenon which characterizes the $s$-dependence of the entropies $\widehat S_m(s)$.
From Eq. \eqref{eq:4.15.v1}, the equipartition is again broken at order $s^2/\mu$.
Anyhow, this subleading term breaking equipartition has not a definite sign with $m$, as an important difference with all considered cases, 
not only in the paper, but in the entire literature. 
This phenomenon can be easily seen by analyzing the difference $\widehat S_m(s)-\widehat S_m(0)$, i.e.,
\begin{equation}
\widehat S_m(s)- \widehat S_m(0)= \frac{(6m- 2^m \sqrt{2^{3-m}+5})}{1-m} \frac{s^2}\mu\,,
\label{corrhat}
\end{equation}
where to get the rhs we explicitly used Eq. \eqref{eq:4.15.v1}.
The coefficient of the term multiplying $s^2$ is negative for $m<m^*=2.695\dots$ and positive for $m>m^*$.
This change of sign causes the R\'enyi entropy to be a monotonous {\it decreasing} function of $|s|$ for $m<m^*$, as all the cases considered so far in the 
literature, while it is a monotonous {\it increasing} function of $|s|$ for $m>m^*$.
It is natural to wonder whether and how this intriguing phenomenon survives to the effect of the further subleading corrections that are not taken 
into account by $g(\s)$ in Eq. \eqref{eq:g}. We will answer this question with the analysis of the numerical data in the next section.

\subsection{The number entropy and the log-log term}

In this subsection we heuristically discuss about the number entropy and its relation with the first subleading term in the symmetry resolved entanglement. 
Eq. \eqref{eq:2.6.a.v1} guarantees that for each realization of the disorder (let us say $r$) it holds
\begin{equation}
S_{r}=-\sum_s p_r(s) \ln p_r(s) + \sum_s p_r(s) S_{r}(s)\,. 
\end{equation}
Taking the average over the disorder means to mediate only {\it after} the sum over $s$ has been performed. 

If we assume {\it self-averaging}, we can invert the two sums/averages, obtaining  
\begin{equation}
\overline S =-\sum_{s} p(s) \log p(s) + \sum_{s} p(s) \overline{S}(s),
\label{self}
\end{equation}
where $p(s)$ is the average probability of configurations with subsystem magnetization $s$ given in Eq. \eqref{eq:4.16.v1}.
Within this assumption, the number entropy is
\begin{equation}
S_{\rm Q}=-\int ds p(s) \ln p(s)= \frac12\Big(1+\ln \frac{\pi \mu}6\Big)\,,
\label{numS}
\end{equation}
i.e., it diverges like $\ln \mu$ for large $\mu$, i.e., like $\ln \ln \ell$. 
This $\ln \mu$ divergence is identical to the one that appears in the symmetry resolved entropy $\overline S(s)$.
Indeed, since at this order in $\mu$ the symmetry resolved entropy $\overline S(s)$ does not show any $s$-dependence, 
we have $\sum_{s} p(s) \overline{S}(s)=\overline S(s)$ and the term $\ln \mu$ (absent in the total entropy $\overline S$) should 
be compensated by an identical term in $\overline S(s)$.

The above equivalence between number entropy and subleading term in the symmetry resolved one takes place in a very similar form also for clean system described 
by a Luttinger liquid. 
There the subleading term $\frac12\ln(K\ln\ell)$ reflects that the charge fluctuations of the subsystem are 
proportional to\cite{bss-07,aem-08,song2010} $K\ln\ell$ (cfr.~\eqref{eq:Zm_CFT}).
The prefactor is again $1/2$ and also cancels in the total entropy when summing number and symmetry resolved ones. 

This is not the end of the story. For clean systems, also the $O(1)$ term in the symmetry resolved entropies is independent of $s$. 
Hence it is also equal to the one for the total entropy (modulo the shift in the number entropy as in \eqref{numS}).
By comparing carefully Eqs. \eqref{eq:4.12.v1}, \eqref{self}, and \eqref{numS} this {\it does not seems to be the case for random systems}. 
Most likely this mismatch is due to the lack of self averaging for the subleading term. 
Another possible explanation could be also the presence of $s$-dependent $O(1)$ terms in the symmetry resolved entropy 
which are not captured by  $g(\s)$ in Eq. \eqref{eq:g} (that, as we stressed, ignores several subleading effects). 

Finally, we have found the same term $-\frac12 \ln \mu$ to be present in {\it all} symmetry resolved entropies independently also of the R\'enyi index. 
This fact can be understood reasoning similarly to what done above.
First,  for $m=1$ it is sufficient to assume self-averaging for the entropy of interest. 
Instead for $m\neq 1$, Eq. \eqref{eq:2.6.a.v1} for the splitting in number and configurational entropy does not hold.
It is also not possible to rewrite a similar form using only the probability $p(s)$.  
However, we can exploit the recent result\cite{crc-20} for a different splitting involving the generalized probabilities $p_m \equiv Z_m(s)/Z_m$ 
($Z_m$ is the exponential of the total R\'enyi entropies).
The complete check is straightforward and not very illuminating, one just needs to assume self averaging for all the quantities of interest. 
In conclusion, this argument explains why the term $-\frac12 \ln \mu$ is present in all the quantities we considered with the same prefactor, 
in spite the coefficient of the leading term (in $\mu=\ln \ell+\dots$) is not the same.

\section{Numerical SDRG results}
\label{sec:num}

In this section we present numerical simulations supporting 
our analytic results. We implement numerically the SDRG method for 
a finite-size Heisenberg chain. 
Specifically, the method works according to the following steps. 
We initialize a list of length $L$ with the chain couplings $J_i$. 
We take the $\{J_i\}$ to be independent random variables with $J_i\in [0,1]$ and extracted from the probability distribution 
\begin{equation}
p(J)= \frac{1}\delta J^{-1+1/\delta} .
\end{equation}
Here $\delta>0$ is a parameter characterizing the strength of the randomness:
$\delta=1$ is the uniform distribution, while $\delta\to\infty$ correspond to strong disorder (i.e., to the RG fixed point). 
We implement the Ma-Dasgupta decimation rule, Eq.~\eqref{eq:3.4.v1}, 
which is iterated on the list of couplings until all the spins are 
decimated. During the iteration the algorithm keeps track of all the singlets that are formed. 
The method is repeated for many random realizations of the couplings. 
From the spatial information about the singlets, it is straightforward 
to calculate the von Neumann and R\'enyi entropies. Given a bipartition of the system 
as $A\cup B$, these are obtained by counting the number of shared singlets 
between the subsystems and by applying~\eqref{eq:vn-nab}. 
The symmetry-resolved entanglement entropies can be calculated in 
a similar way. In fact, in a given disorder realization the SDRG method produces 
$n$ shared singlets. This means that  there are $(n+1)$ blocks. 
Each block, labelled by the quantum number $s$,  is diagonal, and has dimension $d_{s}$ (cf.~\eqref{eq:4.3.v1}). 
Thus, from the spatial configuration of singlets  it is straightforward to 
calculate the symmetry-resolved entropies for each disorder realization and their averages 
$\overline{S}_m(s)$, $\widetilde{S}_m(s)$,  and $\widehat{S}_m(s)$ according to the specific rules in
Eqs.~\eqref{eq:barS}, \eqref{eq:4.5.b.v2} \eqref{Shatav}, respectively.

\begin{figure}[t]
\includegraphics[width=\columnwidth]{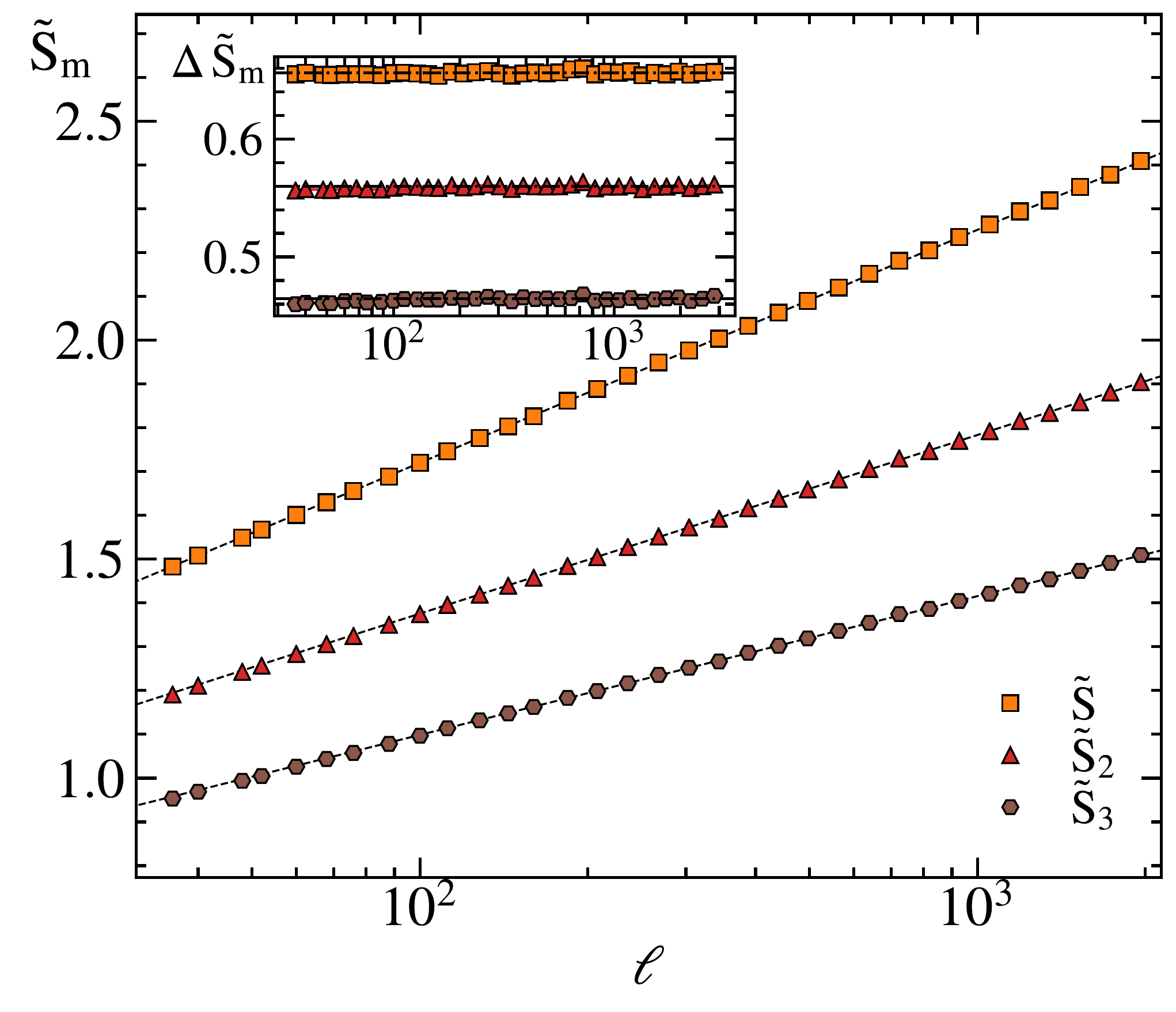}
\caption{Total R\'enyi entropies $\tilde{S}_m$ in the disordered Heisenberg chain, plotted against the subsystem size $\ell$. Symbols represent numerical SDRG simulation with system size ${L=16384}$, disorder strength $\delta=5$, and average over ${N \approx 10^6}$ disorder configurations. 
Different symbols and colors correspond to different R\'enyi index ${m=1, 2, 3}$. Dashed lines are the analytical predictions, Eq.~\eqref{eq:ent-rsp}.
In the inset we report $\Delta \widetilde S_m$ in Eq. \eqref{deltaS} clearly showing that the leading corrections are $O(\ell^0)$.   }
\label{fig:num-pre}
\end{figure}

\subsection{Preliminary benchmarks} 
\label{sec:sdrg-pre}

Before presenting the numerical results for the symmetry-resolved 
entanglement, it is important to reanalyze the behavior of the total, i.e., non symmetry-resolved, von Neumann and R\'enyi entropies. 
In fact, a striking feature of the symmetry-resolved entropies is that they possess subleading double-logarithmic corrections 
that are not present in the total ones. Thus it is worth reanalyzing the total entanglement to exclude log-log terms also here and
to emphasize the differences with the symmetry resolved ones.

Our results for the von Neumann entropy and for the R\'enyi 
entropies $\widetilde{S}_m$ ($m=1,2,3$) are shown in Fig.~\ref{fig:num-pre}.
The symbols are numerical data for a chain with ${L=16384}$ sites (finite size scaling is discussed later) and for a disorder strength $\delta=5$
(other values of $\delta$ provide equivalent result, as discussed for the symmetry resolved ones). 
The data are obtained by averaging over ${N\approx 10^6}$ disorder realizations. 
The continuous lines are the theory predictions obtained as a fit of the form 
\begin{equation}
\label{eq:ent-rsp}
	\widetilde{S}_m = 	
	\frac{\sqrt{5+2^{3-m}}-3}{2(1-m)}\ln\ell+a, 
\end{equation}
in which $a$ is the only fitting parameter. 
The agreement between the SDRG results and~\eqref{eq:ent-rsp} is good for all the entropies.  
However, in order to have a better feeling of the subleading term, in the inset we plot the subtracted entropy
\begin{equation}
\Delta \widetilde S_m\equiv \widetilde{S}_m -\frac{\sqrt{5+2^{3-m}}-3}{2(1-m)}\ln\ell\,.
	\label{deltaS}
\end{equation}
This inset provides a strong evidence that the leading correction to the entropy is $O(\ell^0)$, ruling out the presence of a log-log term.

We do not report the numerics for $\overline{S}_m$ because they coincide with $\widetilde{S} =\widetilde{S}_1$ by definition (see the 
discussion in Section~\ref{sec:ent-rsp}).

\begin{figure}[t]
\includegraphics[width=\columnwidth]{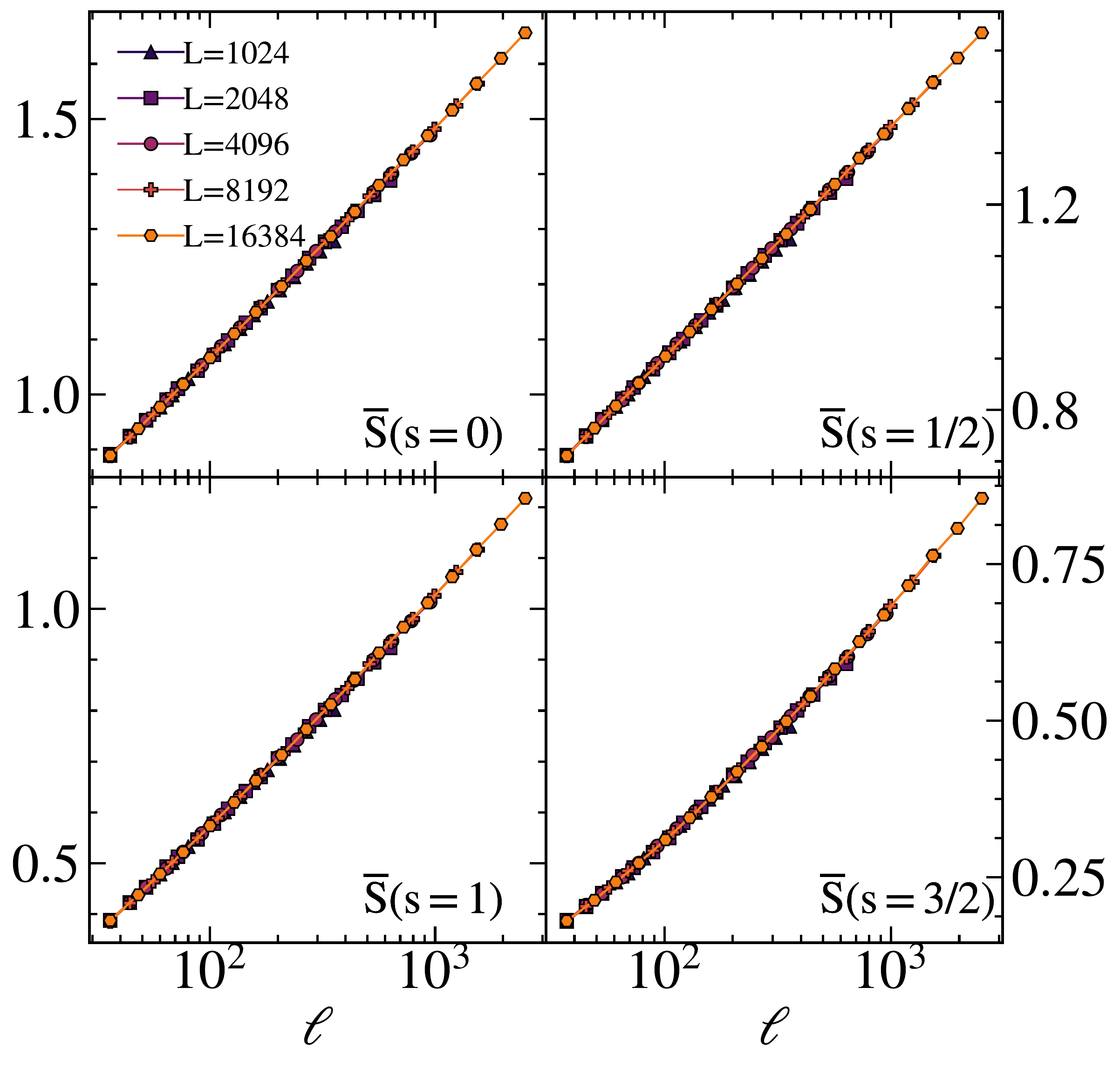}
\caption{Finite size behavior of the disorder averaged symmetry resolved entropies $\overline S(s)$.
We report the numerical data for $L$ from $1024$ to $16384$. The four panels are for $s=0,1/2,1,3/2$.
All data correspond to disorder strength $\delta=1$ (uniform distribution) and are averaged over $10^6$ disorder realizations. 
For the considered values of $\ell$ there are no visible finite size corrections.
 }
\label{fig:fss}
\end{figure}

\subsection{Symmetry-resolved von Neumann entropy} 
\label{sec:sdrg-vN}

We now discuss the symmetry-resolved von Neumann entropy in the random singlet phase. 
We  will consider both $\overline S(s)$ in Eq.~\eqref{eq:Srsp_numerics} and $\widehat{S} (s)$ in Eq.~\eqref{hatS_vN}.
We recall that $\overline S(s)$ is the limit ${m\to1}$ of both $\overline{S}_m(s)$ and $\widetilde{S}_m(s)$.

We start with the analysis of the finite size behavior. 
In Fig. \ref{fig:fss} we report the numerical data for $\overline S(s)$ at fixed disorder $\delta=1$ for $L=1024,2048,4096,8192,16384$.
The averages are over ${\approx 10^6}$ disorder realizations. 
The subsystem magnetization $s$ can assume both integer of semi-integer values, depending on the parity of $\ell$.
Hence, hereafter the data for $s=0,1,2$ correspond to even $\ell$, while the data for $s=1/2, 3/2$ are for odd $\ell$
(everywhere for each even $\ell$ considered, we also plot $\ell+1$). 
In Fig. \ref{fig:fss}, the data for all reported values of $\ell$ are on top of each other and there is no visible finite size correction for any $s$.
Then in the following, we will work mainly at $L=16384$ and consider values of $\ell$ up to $\sim 3000$ for which there are no appreciable corrections. 
We checked that this feature is universal, i.e., does not depend either on disorder strength $\delta$ or on the considered entropy. 
In all this paper, we only report positive values of $s$, but we tested that for $s\to-s$ we get exactly the same results.

\begin{figure}[t]
\includegraphics[width=\columnwidth]{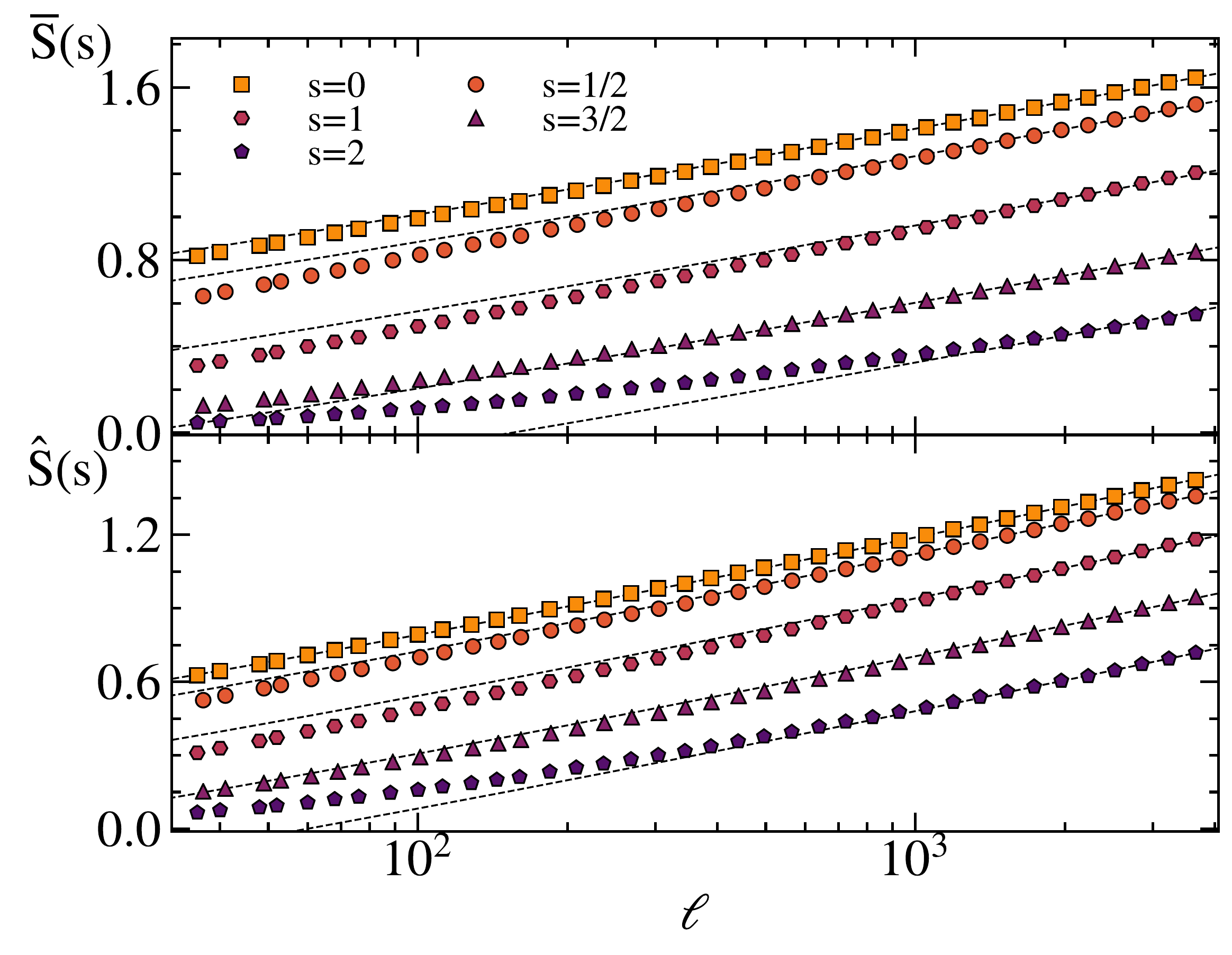}
\caption{Symmetry resolved entanglement entropy $\overline S(s)$ (top) and $\widehat{S}(s)$ (bottom) in the disordered Heisenberg chain, against the subsystem size $\ell$. 
Symbols are numerical results using the SDRG method with system size ${L=16384}$ and average over ${N\approx 10^6}$ disorder realizations with disorder strength $\delta=5$. 
Different symbols and colors correspond to different symmetry sectors  ${s=0,1/2, 1,3/2,2}$. 
The  dashed lines represent the theory prediction~\eqref{eq:fits-vN1} in which $a$ has been adjusted to fit the data. 
 }
\label{fig:tildeSvN_symm}
\end{figure}

We are now ready to start our analysis of the symmetry resolved entropies. 
In Fig.~\ref{fig:tildeSvN_symm} we compare our analytical formulas with numerical SDRG results.
The data are for a chain with ${L=16384}$ sites and are obtained by averaging over ${N \approx 10^6}$ disorder realizations. 
Data are plotted as function of $\ell$. 
The different symbols correspond to the different symmetry sectors $s$.
We report the data for both $\overline S(s)$ (top panel) and $\widehat S(s)$ (bottom panel).  
For large $\ell$ all curves become parallel, showing asymptotic equipartition, as we theoretically derived in the previous section. 
The curves however are not superimposed, manifesting that the subleading corrections do depend on $s$. 
Furthermore we find that the various curves are monotonously decreasing function of $|s|$, as theoretically predicted in the previous section. 
In this respect, it is important that non-universal terms not included in the approximations that led to Eqs. 
\eqref{corrbar} and \eqref{corrhat} do not spoil such a result.

%

\begin{figure}[t]
\includegraphics[width=\columnwidth]{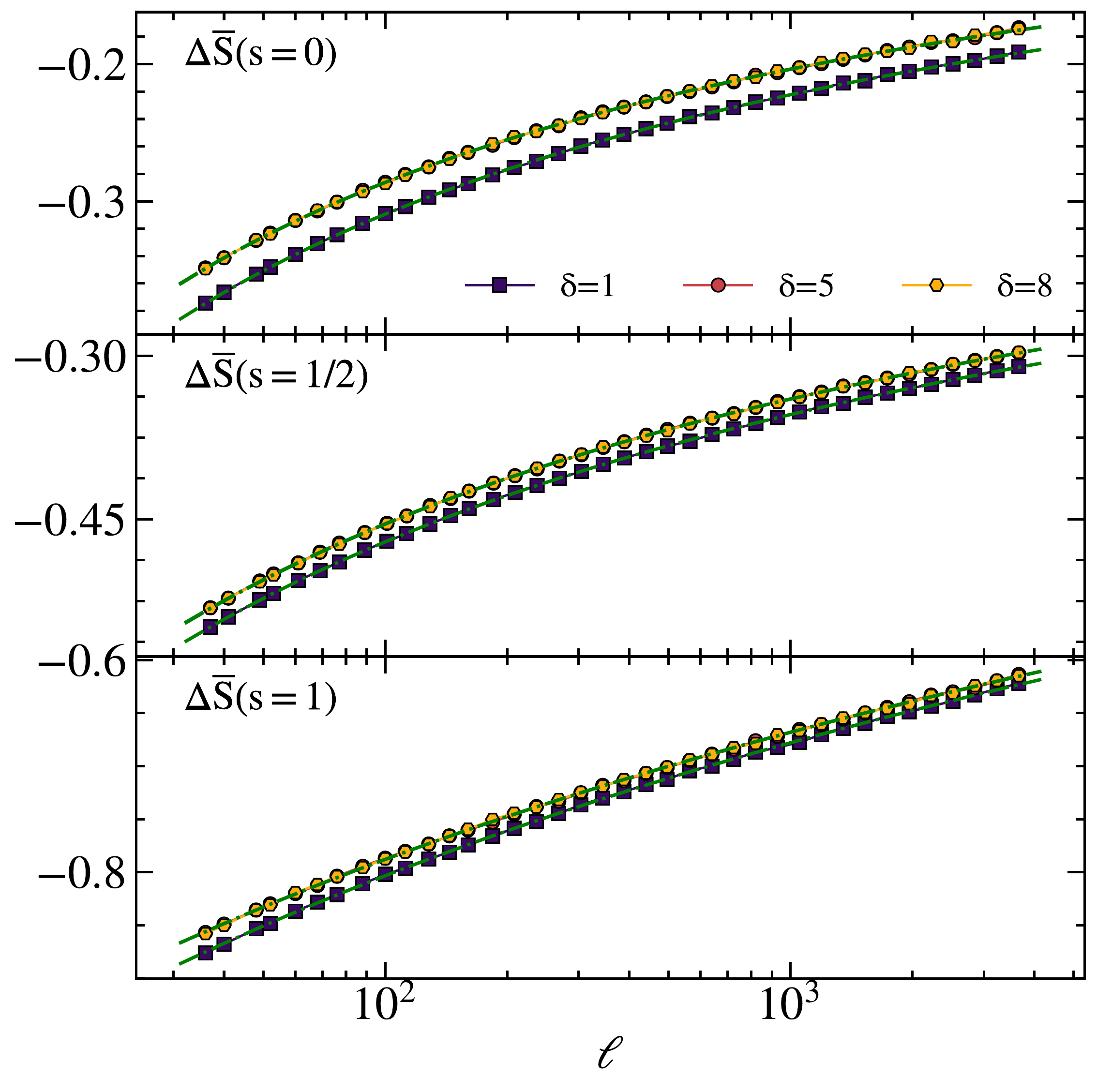}
\caption{Subtracted symmetry resolved entanglement entropy $\Delta S (s) $ in Eq. \eqref{DeltaSs} for $s=0,1/2,1$ (from top to bottom) in the disordered Heisenberg chain, 
against the subsystem size $\ell$. Symbols are numerical results using the SDRG method with system size ${L=16384}$ and average over ${N\approx 10^6}$ disorder realizations. Different symbols and colors correspond to different disorder strength $\delta=1,5,8$.
The data are slowly approaching $0$ according to the law \eqref{Dsfit}. 
The fits are reported as lines and perfectly match the data. 
 }
\label{fig:corr}
\end{figure}

The dashed  lines in figure \ref{fig:tildeSvN_symm} are fit to the form
\begin{equation}
\label{eq:fits-vN1}
\frac{\ln(2)}{3}\ln\ell- \frac{1}{2}\ln\ln\ell +a,
\end{equation}
with a single free parameter $a$. We use the same form for both $\overline S(s)$ and $\widehat S(s)$ since in SDRG they show the same asymptotic 
scaling (cf. Eq. \eqref{eq:4.17.v1}) with a different $O(\ell^0)$ term, i.e., with a different $a$ in the above equation. 
The fit is performed only with the data for large $\ell$. 
The agreement is really good taking into account that we only have one parameter in the fit. 
It is clear that the corrections to the scaling become more important for larger values of $|s|$, as it was expected on the 
bases of the result of the previous section. 
Needless to say that the presence of the term $-1/2 \ln\ln \ell $ in Eq. \eqref{eq:fits-vN1} is fundamental to have such agreement.

However, proceeding in this way, we would have the additive constant $a$ which does depend on $s$.
Conversely, the SDRG results in Eqs. \eqref{corrbar} and \eqref{corrhat} suggest that this is not the case.
It is also true that our SDRG treatment ignores some non-universal processes that do not alter the two leading terms, but at least in principle can affect the constant. 
On the other hand, within SDRG we have also shown the presence of $s$-dependent terms behaving like $s^2/\ln\ell$
(indeed these log-corrections are typical features of symmetry resolved entanglement entropies also in clean systems~\cite{Bonsignori2019,crc-20}). 
Can these corrections be responsible for a seemingly $s$-dependent additive constant? 
To answer this question we study the difference
\begin{equation}
\Delta \overline S(s)\equiv \overline S(s)- \overline S+ \frac{1}{2}\ln\Big(\frac{\pi}6\ln\ell \Big)\,.
\label{DeltaSs}
\end{equation}
This subtraction is motivated by the fact that the additive constant not only is $s$ independent, but also equal the one for the total entropy (modulo 
the additive factor within the number entropy). 
Hence, according to our SDRG results, $\Delta \overline S(s)$ should decay to zero for large $\ell$ as
\begin{equation}
\Delta \overline S(s)\simeq \frac{b}{\ln \ell+c}\,,
\label{Dsfit}
\end{equation}
where $b$ and $c$ are free non-universal parameter that may (and actually do) depend on $s$.
We analyze the SDRG data for $\Delta \overline S(s)$ in Fig. \ref{fig:corr} where we consider
three different disorder distributions with strength $\delta=1,5,8$ to rule out the possibility of some weak disorder effect. 
It is evident that $\delta$ only mildly influences the data and for $\delta=5$ and $\delta=8$ there are no differences at all. 
The three panels in the figure correspond to $s=0,1/2,1$.
The numerical data are fit to the form \eqref{Dsfit}. The agreement is truly impressive when one thinks that we 
are fitting curves that asymptotically tend to zero, but we are working in a regime where they are still far from it. 
Increasing the values of $s$, subleading terms, e.g. going like $s^2/(\ln\ell)^2$ or $s^4/(\ln\ell)^\alpha$, becomes important and it is more difficult to fit the data with \eqref{Dsfit} at the available values of $\ell$. 
Finally we mention that we repeated the same analysis also for the entropy $\widehat S(s)$ finding equivalent results for the equipartition.

Concluding, Fig. \ref{fig:corr} is a very strong and convincing evidence that the prediction from SDRG in Eq. \eqref{corrbar}
survives the inclusion of non-universal effects and that there is equipartition of entanglement at the order $O(\ell^0)$ also in 
the random singlet phase. The first term breaking equipartition $s^2/\ln\ell$ is also correctly captured by SDRG in Eq. \eqref{corrbar}.

\subsection{Symmetry-resolved R\'enyi entropies}
\label{sec:sdrg-renyi}

\begin{figure}[t]
\includegraphics[width=\columnwidth]{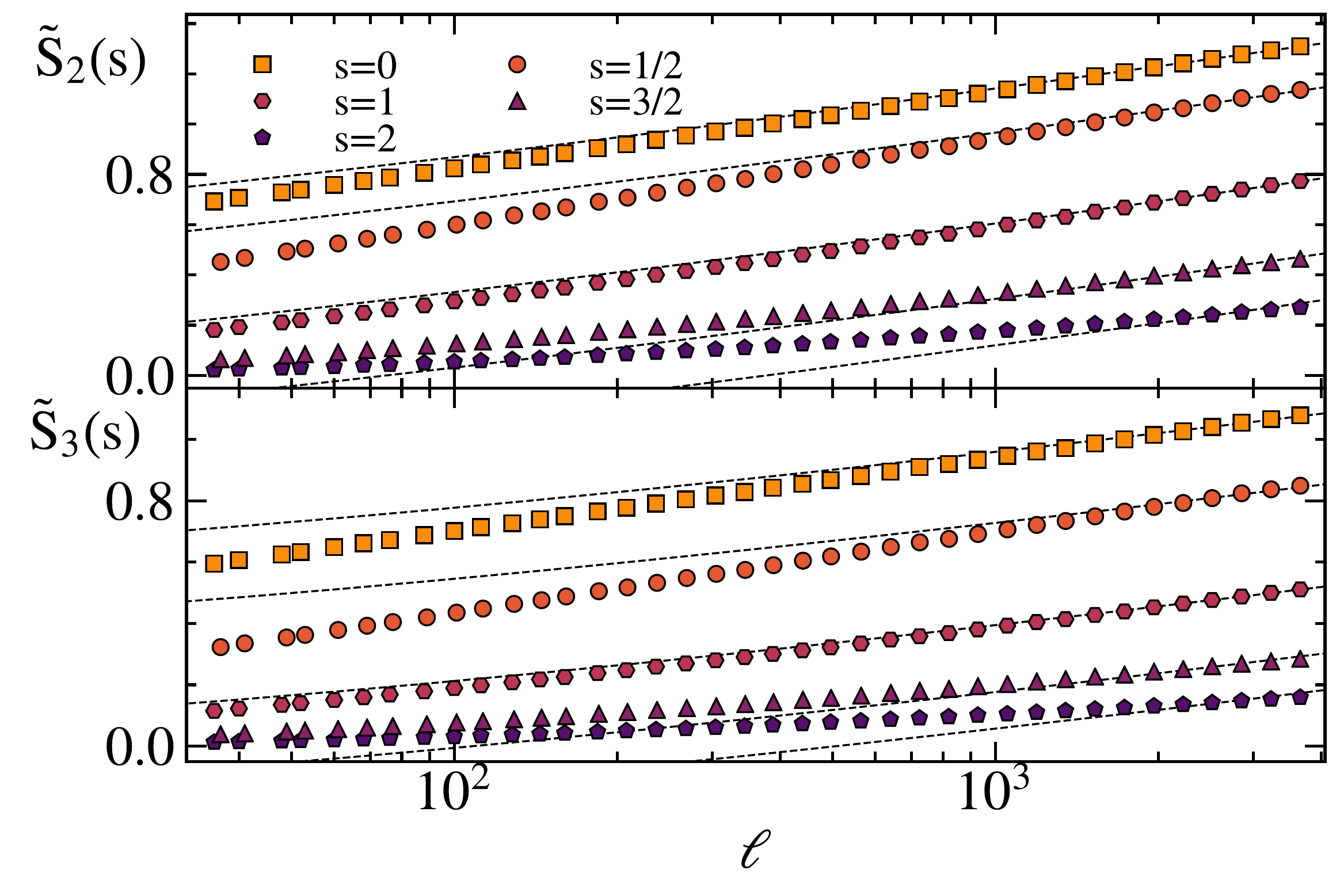}
\caption{Symmetry-resolved R\'enyi entropies $\widetilde{S}_m(s)$ for $m=2$ (top) and $m=3$ (bottom). 
 The symbols are SDRG results for the Heisenberg chain with  ${L=16384}$ sites. The average is over $10^6$ disorder realizations with strength $\delta=5$. 
 Different symbols and colors correspond to different subsystem magnetization ${s=0, 1/2, 1,3/2,1}$.
 The dashed lines are fits to the form~\eqref{eq:fits-r}.
}
\label{fig:2.v1}
\end{figure}

We now discuss the symmetry-resolved R\'enyi entropies. 
For many aspects the analysis is identical to the one of the previous section for the von Neumann one and we will not repeat all details.

We first consider $\widetilde{S}_m (s)$ (since $\overline{S}_m(s)$ do not depend on $m$, there is no reason to discuss them). 
In Fig.~\ref{fig:2.v1} we plot SDRG data for $m=2,3$. 
In the scaling limit all the R\'enyi entropies exhibit equipartition and are described by 
\begin{equation}
\label{eq:fits-r}
\widetilde{S}_m(s)=
\frac{\sqrt{2^{3-m}+5}-3}{2(1-m)} \ln\ell - \frac{1}{2}
\ln\ln\ell+ a+\dots. 
\end{equation}
Again, the first term in~\eqref{eq:fits-r} is the result for 
the total R\'enyi entropies, $\widetilde{S}_m$. Note that the subleading term 
$1/2\ln\ln\ell$ is the same as for the von Neumann entropy. 
For large $\ell$ all curves at fixed $m$ become parallel, showing asymptotic equipartition. 
Anyhow, they are not on top of each other, manifesting that the subleading corrections do depend on $s$.

The first check to test the asymptotic behavior is to perform a simple fit of the data to the form \eqref{eq:fits-r}
allowing $a$ to depend on $s$. These fits are shown in Fig.~\ref{fig:2.v1} as continuous lines. The agreement is excellent
and, as expected, it slowly deteriorates increasing $|s|$.
We have performed an analysis like the one in Fig. \ref{fig:corr} for the von Neumann entropy to convince ourselves 
that the differences between the various curves at fixed $s$ are, as SDRG predicts in Eq. \eqref{corrtil}, only due to subleading term as $s^2/\ln \ell$.
The analysis shows that this is likely, but the corrections are much larger than for $m=1$ and so more difficult to treat. 
This is not unexpected: Eq. \eqref{corrtil} predicts that the coefficient of $s^2/\ln \ell$ grows exponentially with $m$ 
(being $\sim 6$ at $m=1$, $\sim10.6$ at $m=2$, and $\sim 20$ at $m=3$) and so the data soon become difficult to handle as $m$ increases.

Finally, we discuss the R\'enyi entropies $\widehat{S}_m(s)$, focusing on ${m=2, 3}$. 
Before discussing the scaling behavior of the entanglement entropies, it is useful to consider the partition functions
$\langle {Z}_m (s)\rangle $, being the main ingredient to construct $\widehat{S}_m(s)$ and being per se interesting 
(for $m=1$, $\langle {Z}_1 (s)\rangle$ is the average probability $p(s)$ of having subsystem magnetisation $s$, cf. Eq. \eqref{eq:4.16.v1},
while for other $m$ are related to generalized probability distributions\cite{crc-20}).

\begin{figure}[t]
\includegraphics[width=\columnwidth]{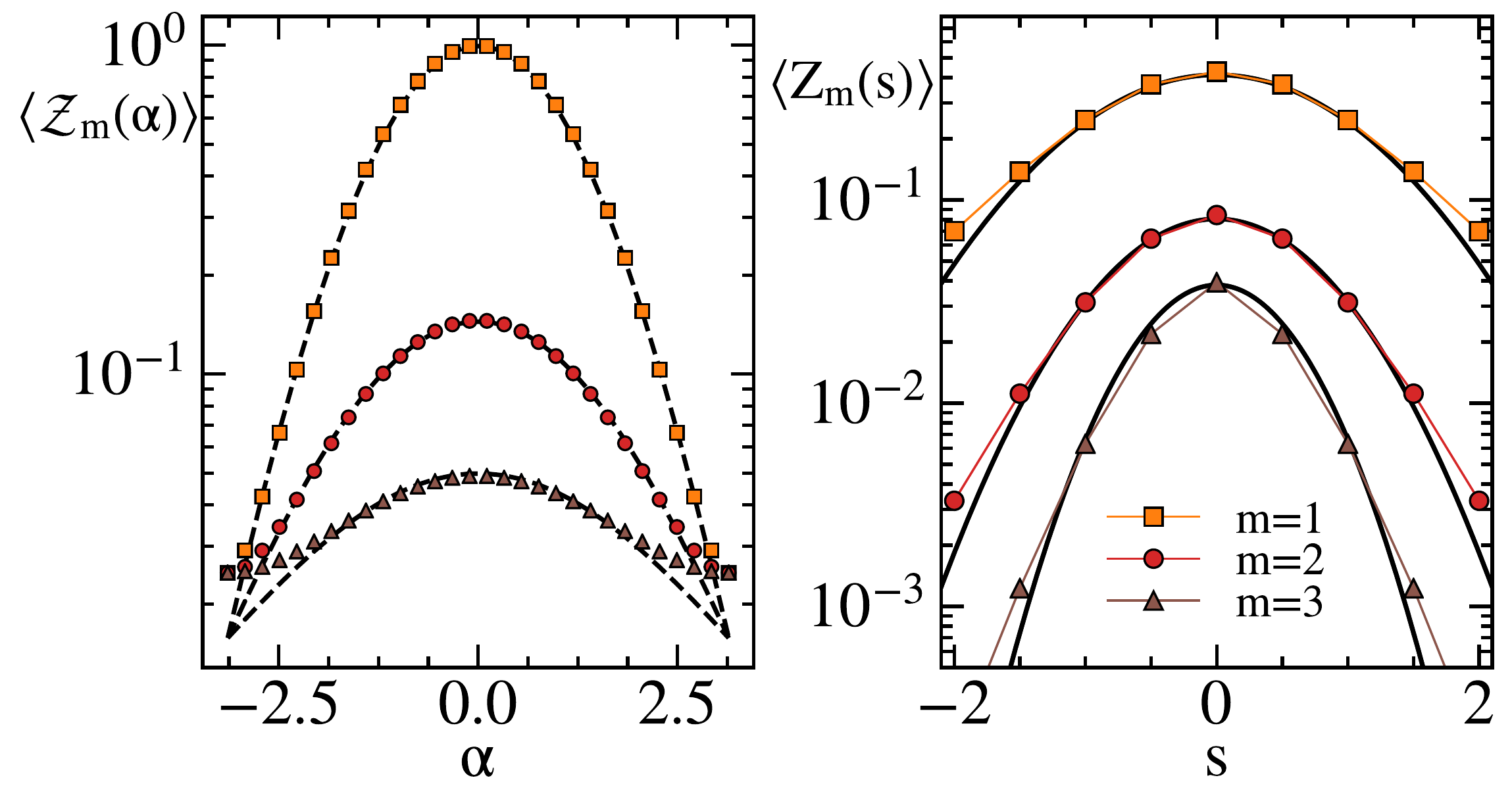}
\caption{Charged (left) and symmetry-resolved (right) moments, $\langle \mathcal{Z}_m (\alpha)\rangle $ and $\langle Z_m (s) \rangle$.
Symbols represent numerical SDRG simulation with system and subsystem sizes ${L=8192}$ and ${\ell = 3722}$, respectively. 
The average is over $N \approx10^6$ disorder configurations and the strength of the disorder is $\delta=1$. 
Different symbols and colors correspond to different index ${m=1,2, 3}$. 
Dashed and full lines are the analytic predictions, Eq.~\eqref{eq:4.14.v1} and Eq.~\eqref{eq:4.15.v1}, respectively. 
\label{fig:Zm_alpha_sq}}
\end{figure}

In Fig.~\ref{fig:Zm_alpha_sq}, we present a quantitative comparison for $\langle {\mathcal{Z}}_m(\alpha)\rangle$ and  $\langle Z_m (s) \rangle$ between the  numerics 
and the analytic predictions, respectively in Eqs.~\eqref{eq:4.14.v1} and ~\eqref{eq:4.15.v1}. 
The additive constant in $\mu$ (cfr. Eq.~\eqref{eq:mu}) appearing in both formulas is preliminary fitted only once for all data. 
The numerical data are obtained by averaging Eq.~\eqref{eq:4.13.v1} for $\langle {\mathcal{Z}}_m(\alpha)\rangle$ 
and Eq. \eqref{eq:4.4.a.v1} for  $\langle Z_m (s) \rangle$.
For $\langle \mathcal{Z}_m(\alpha)\rangle $, we observe a fair agreement between our data and the analytic expressions, 
although finite size corrections are present as $\alpha\to\pm\pi$ (the plot is in log scale). 
The discrepancies at the boundaries of the Brillouin zone are well known for clean systems \cite{Bonsignori2019,Fraenkel2019,sara2D}
and are physically due to the fact that such charged entropies must be periodic of period $2\pi$.
Then they should be present in lattice disorder systems as well. 
Also the data for  $\langle Z_m(s)\rangle$ are remarkably reproduced by SDRG predictions, with corrections to the scaling that become
larger as $|s|$ increases, as it is the case for all the quantities considered so far. 
Incidentally, we did not yet mention that, very generically, deviations at higher $s$ are expected, because populating higher sectors requires exponentially larger system sizes.

\begin{figure}[t]
\includegraphics[width=\columnwidth]{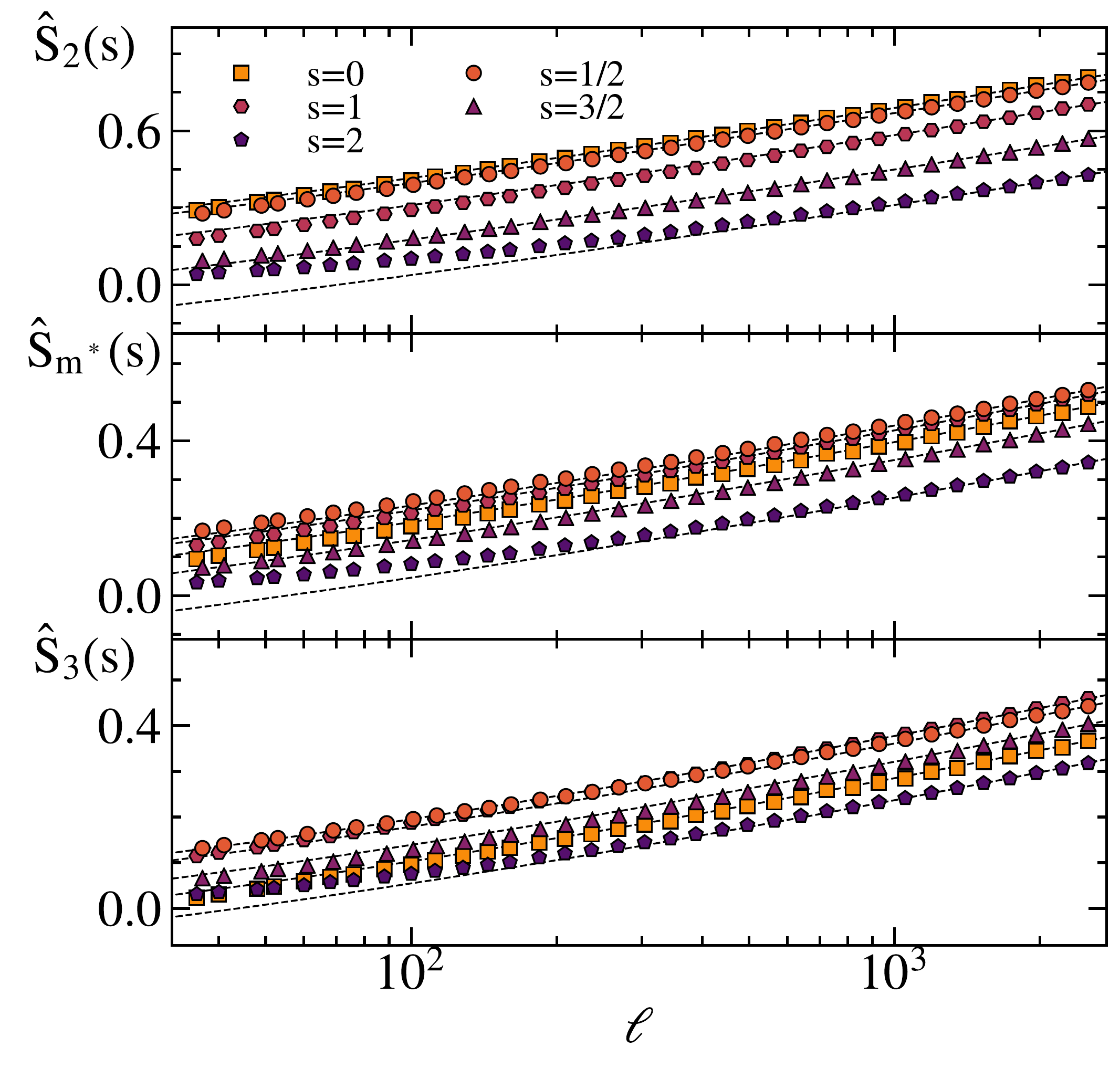}
\caption{Symmetry-resolved R\'enyi entropies $\widehat{S}_m(s)$ for $m=2$, $m=m^*=2.69\dots$, and $m=3$ (from top to bottom). 
The symbols are SDRG results for the Heisenberg chain with  $L=16384$ sites and disorder strength $\delta=5$. The average is over $10^6$ disorder realizations. 
Different symbols and colors correspond to different subsystem magnetization $s=0,1/2, 1$.
The lines are fits to the theoretical result~\eqref{eq:fits-r3}.
Notice that the entropies are ordered as monotonically decreasing function of $s$ for $m=2$,  are mixed up at $m=m^*$, and start inverting their order  
for $m=3$.%
	\label{fig:hatS_m_sq}}
\end{figure}

We are then ready to analyze the symmetry-resolved entropies $\widehat{S}_m(s)$ which are plotted in Fig.~\ref{fig:hatS_m_sq}. 
Again, for asymptotic large $\ell$ the various curves for different $s$ at fixed $m$ become parallel, manifesting equipartition. 
As done for all other entropies, we first check the correctness of the leading scaling term that in SDRG is given by Eq. \eqref{eq:4.17.v1}.  
In Fig.~\ref{fig:hatS_m_sq}, the continuous lines are fit of the data with
\begin{equation}
\label{eq:fits-r3}
\widehat{S}_m=\frac{\sqrt{2^{3-m}+5}-3}{2(1-m)} \ln\ell - \frac{1}{2}
\ln\ln\ell+ a+\dots  
\end{equation}
where we allow $a$ to depend on $s$. We observe a good asymptotic agreement in Fig.~\ref{fig:hatS_m_sq} confirming the correctness of the leading term. 

We now move to the corrections. In this case the analysis is very difficult because of the peculiar non-monotonic features we highlighted at the end 
of Sec. \ref{sec:sbar} in Eq. \eqref{corrhat}. Indeed while they are always of the form $s^2/\ln\ell$, the prefactor is negative for $m<m^*$ (as in all other 
cases observed so far here and in the literature) and it is positive for $m>m^*$. 
Consequently the entropies are expected to be monotonous decreasing functions of $|s|$ for $m\alt m*$ and monotonous increasing function of $|s|$ for $m\agt m*$.
Close to $m^*$, the subsubleading terms become important and larger than the ones under scrutiny that instead vanish at $m=m^*$.  
Exactly for this reason in the Fig.~\ref{fig:hatS_m_sq} we report $m=2$, $m=m^*$, and $m=3$. 
We observe that the entropies are ordered as monotonically decreasing function of $s$ for $m=2$, 
they are mixed up at $m=m^*$ (which is the point where the leading corrections to the scaling vanish in SDRG) and 
they tend to reverse their order for $m=3$, although they are not in increasing order in $s$, likely because of subleading corrections ($m^*$ is very close to 3).  
We found extremely remarkable that this unusual effect predicted by SDRG is not spoiled by non-universal effects as well as by other universal RG processes
that have not been included in the derivation of $g(\s)$ in Eq. \eqref{eq:g} presented in Ref. \onlinecite{Fagotti2011}.

\section{Discussion}
In this paper we investigated the symmetry resolved entanglement in the random singlet phase. 
Because of the average over random disorder, we have three possible alternative definitions of symmetry resolved R\'enyi entropies 
that we give in Eqs. \eqref{eq:4.5.a.v1}, \eqref{eq:4.5.b.v1}, and \eqref{eq:4.5.c.v1}. Two of them (\eqref{eq:4.5.a.v1} and \eqref{eq:4.5.b.v1}) 
become equal in the von Neumann limit.
We compute the asymptotic behavior of these entropies in the large $\ell$ limit using 
well established techniques within SDRG. 
Our main result is that the three definitions all provide entanglement entropies that satisfy equipartition at the leading universal orders. 
We confirmed these results numerically and showed the presence of subleading non-universal terms breaking equipartition.
The order of such corrections, $s^2/\ln\ell$, is also correctly characterized by analytic SDRG techniques. 
We finally point out that the double logarithmic term in the symmetry resolved entanglement is related to the number entropy, 
in full analogy with clean systems \cite{Bonsignori2019}.
There are also few quantitative remarkable SDRG predictions about the subleading terms that are confirmed by numerics. 
The first is that the $O(\ell^0)$ term in the symmetry resolved entanglement, not only is $s$-independent (a remarkable fact by itself),
but it is also the same as in the total entropy (modulo a contribution from the number entropy). 
Another one is that for almost all entropies the corrections are monotonically decreasing function of $|s|$, but for the one defined in 
Eq. \eqref{eq:4.5.c.v1} there is a switch as the R\'enyi index grows. 

An important test of our results that is still to be performed consists in checking some of our predictions  in microscopic models with ab-initio methods. 
However, it is a numerically demanding problem to reach the large system sizes required to minimize the effect of the subleading corrections, 
even for disordered free-fermion models and exploiting well established techniques\cite{Laflorencie2005,Fagotti2011}.

A fundamental generalization of our work concerns symmetry resolved entanglement and equipartition for disordered systems out of equilibrium. 
Indeed, there is already a large literature about the time evolution of the total entanglement entropy \cite{zpp-07,Dechiara2006,Bardarson2012,Serbyn13,Vosk2014,Altman15,Parameswaran2017,Abanin18,Pekker2014,Zhao2016,Lukin2019,exp-mbl}, 
that provided insights also about the celebrated many body localization. 
Only in recent experiments\cite{Lukin2019}, the importance of symmetry resolution has been highlighted also to shed light on the slow growth of 
the total entanglement entropy. However, many aspects of the problem still require to be studied deeply.

\acknowledgments 
PC acknowledges support from ERC under Consolidator grant  number 771536 (NEMO).
VA acknowledges support from ERC under Advanced grant 743032 (DYNAMINT).
XT acknowledges support from ERC under Starting grant 758329 (AGEnTh).
\appendix

\section{Moments of the size of the symmetry block $I_m$}
\label{app:comp}

In this appendix we show how to use the Laplace transform techniques to rigorously calculate the leading behavior of $I_m(s)$. 
First of all we notice that for large $n$, the sum in Eq.~\eqref{eq:4.6.v1} can be replaced by an 
integral, and we can exploit the  closed-form expression for the 
generating function $g(\s)$ of the moments of $n$ (cf. Eq.~\eqref{eq:3.11.v1}), which is
the Laplace transform of the probability distribution, i.e., ${g(\s) = \mathcal{L}_n [P(n)](\s)}$. 

Thus, to evaluate Eq.~\eqref{eq:4.6.v1}, we first introduce the following Laplace transform
\begin{equation}
\label{eq:4.7.a.v1}
f(\s;s) \equiv \mathcal{L}_n\left[\binom{n}{n/2+ s}^{(1-m)}\right].
\end{equation}
Then, by using the rule for the Laplace transform of a product, Eq.~\eqref{eq:4.6.v1} can be written,  in the scaling regime of large $n$, as 
\begin{equation}
\label{eq:integral}
I_m(s) = \lim_{T\to \infty} \frac{1}{2\pi i} \int_{a-iT}^{a+iT}  g(\s) f(-\s;s) d\s,
\end{equation}
where $a$ is a real number that guarantees convergence of the integral
and $g(\s)$ is the generating function of the moments of the 
distribution of the number of shared singlets defined in Eq.~\eqref{eq:3.11.v1}. 

The Laplace transform of~\eqref{eq:expansion0} with respect to $n$ 
can be calculated order by order by using that 
\begin{multline}
\label{eq:laplace}
{\mathcal L}_\s(2^{n(1-m)}n^{\frac{m}{2}}n^{-2k-\frac{1}{2}}n^\beta)=\\
\Gamma\Big(\frac{1+m}2-2k+\beta\Big)(\s+(m-1)\ln2)^{2k-\beta-\frac{m+1}2}.
\end{multline}
We now observe that the generating function $g(s)$ is analytic in the complex plane. 
On the other hand, Eq.~\eqref{eq:laplace} shows 
that if $m$ is even $f(-\s;s)$ has an algebraic  branch point  
at $\s^*=(m-1)\ln2$. Instead, for $m$ odd there is a pole at $\s^*$ when $2k-\beta-(m+1)/2<0$, 
which is the reason why it becomes just a derivative, as in the main text.

\begin{figure}
\includegraphics[width=.6\columnwidth]{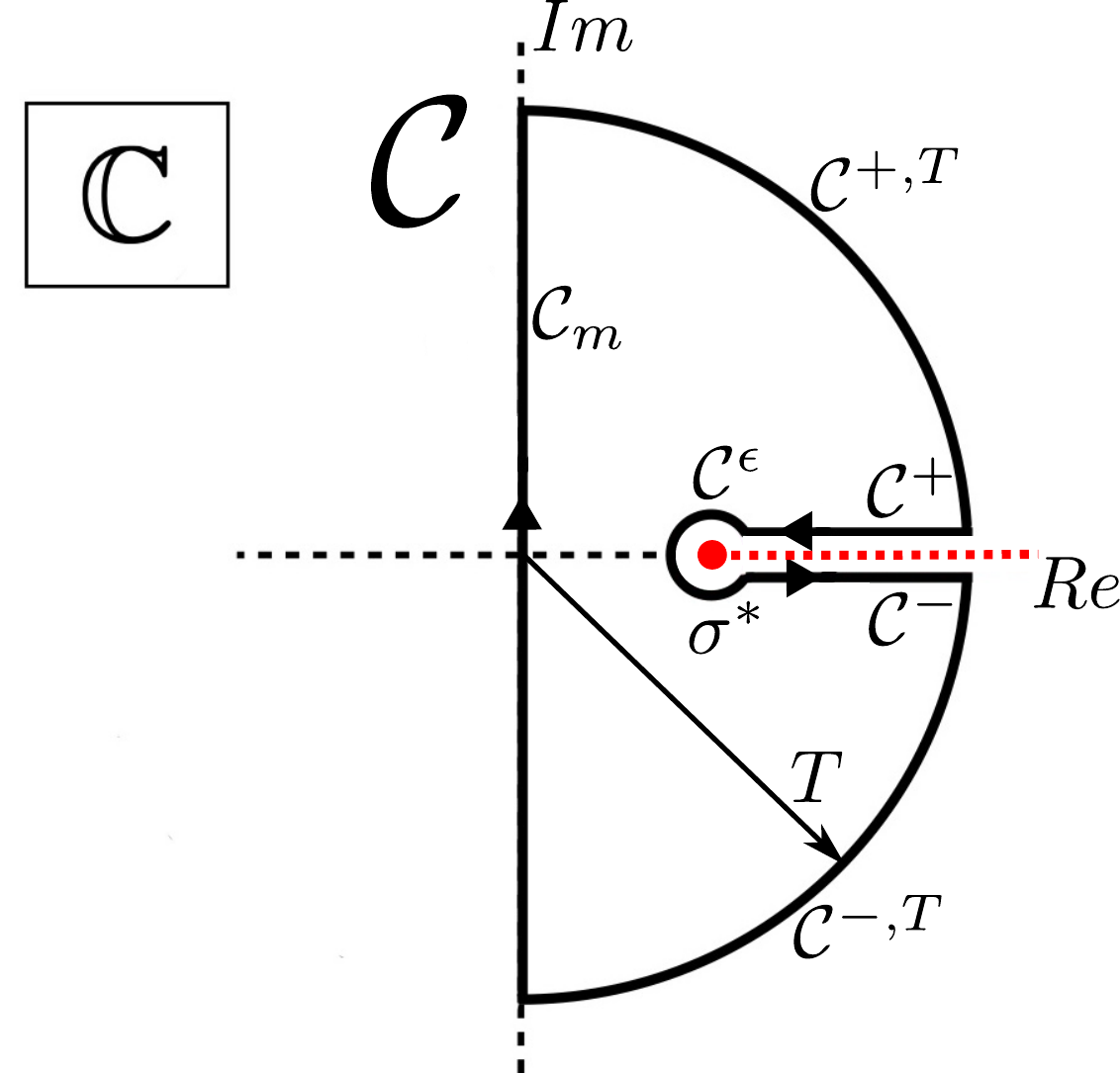}
\caption{Contour ${\mathcal C}$ to evaluate the integral in 
 Eq.~\eqref{eq:integral}. The integral on ${\mathcal C}_m$ is 
 \eqref{eq:integral}. For even values of the R\'enyi index 
 $m$ there is a branch cut starting at $\s^*=(m-1)\ln(2)$. For odd 
 $m$ the contributions of ${\mathcal C}^\pm$ cancel out and one has 
 a pole at $\s^*$. The contributions of the semicircle $C^{\pm,T}$ vanishes.
}
\label{fig:app}
\end{figure}

The integral~\eqref{eq:integral} can be performed by considering the 
contour integration along the path in Fig.~\ref{fig:app} (we report $m$ even, for $m$ odd it is slightly simpler). 
The red dashed line is the branch cut starting at the algebraic branch point singularity at  
$\s^*=(m-1)\ln2$. Since there are no singularities in the region enclosed by the contour, the integral is zero, 
i.e., 
\begin{multline}
\label{eq:contour}
\frac{1}{2\pi i}\int_\mathcal{C} g(\s) f(-\s;s) d\s= \\
I_m + I^+ + I^- + I^{+,T} +I^{-,T}+ I^\epsilon=0.
\end{multline}
Here $I_m$ is the integral~\eqref{eq:integral}, where we set $a=0$, 
$I^\pm$ are the integrals on the paths ${\mathcal C}^\pm$, 
${\mathcal C}^{\pm,T}$ are contributions of the large semicircle, 
and $I^\epsilon$ is the contribution of ${\mathcal C}^\epsilon$. 

%
We are interested in the limit $T\to\infty$ ($T$ is the radius of the semicircle). 
It is straightforward to show that in this limit the contribution of $I^{+,T}+I^{-,T}$ vanishes. 
We should also observe that for odd $m$ the 
two terms $I^\pm$ cancel out because the singularity at $\s^*$ is a pole. 
Here, not to loose generality, we consider the case of $m$ even, while odd can be deduced as a special case. 

From Eqs.~\eqref{eq:integral} and~\eqref{eq:laplace}, 
the integrals in~\eqref{eq:contour} are of the form 
\begin{equation}
\label{eq:int-ex}
{\mathcal I}_{{\mathcal C}'}=\frac{1}{2\pi i}\int_{\mathcal{C}'} 
\frac{g(u-\s^*)}{(-u)^{\omega}}du ,
\end{equation}
where ${\mathcal C}'$ denotes the different paths forming the 
contour in Fig.~\ref{fig:app}, and we defined 
\begin{align}
\label{eq:def-1}
& u\equiv \s-\s^*,\\
\label{eq:def-2}
& \omega\equiv (m+1)/2-2k+\beta.
\end{align}
Here we are interested in the limit $\ell\to\infty$, which implies 
$\mu\to\infty$ (cf.~\eqref{eq:mu}). In this limit we can simplify the 
expression for the generating function $g(\s)$ as 
\begin{equation}
\label{eq:genlarge}
g(\s)\simeq \frac{1}{2}e^{-\frac{1}{2}(3-\sqrt{5+4e^{-\s}})\mu}
\Big[1+\frac{3}{\sqrt{5+4e^{-\s}}}\Big]. 
\end{equation}
The integral~\eqref{eq:int-ex} is difficult to compute in general. 
However, in the limit $\mu\to\infty$ on can use the saddle point method. 
Let us start discussing the contribution of the path ${\mathcal C}^+$:
\begin{multline}
{\mathcal I}_{{\mathcal C}^+}=-\frac{1}{2\pi i}\int_\epsilon^{T'}
\frac{e^{-\frac{1}{2}(3-\sqrt{5+2^{3-m}e^{-u}})\mu}}{2(-u)^{\omega}}
\\ \times \Big[1+
\frac{3}{\sqrt{5+2^{3-m}e^{-u}}}\Big]du, 
\end{multline}
where $T'=T-\s^*$.
A standard saddle point analysis of this integrals in the large 
$\mu$ limit gives the leading contribution as 
\begin{multline}
\label{eq:s1}
{\mathcal I}_{{\mathcal C}^+}^{(0)}=
\frac{e^{-\frac{1}{2}(3-\gamma_m)\mu}}
{4\pi i(-1)^\omega}\Big(1+\frac{3}{\gamma_m}\Big)\\
\times\Big(\frac{2^{1-m}\mu}{\gamma_m}\Big)^{\omega-1}
\left.\Gamma\Big(1-\omega,
\frac{2^{1-m}u}{\gamma_m}\Big)\right|_{\mu\epsilon}^{\mu T'}, 
\end{multline}
where $\Gamma$ is the incomplete Gamma function and we introduced
\begin{equation}
\gamma_m=\sqrt{5+2^{3-m}}.
\end{equation}

Saddle point corrections to Eq.~\eqref{eq:s1} are ${\mathcal O}(1/\sqrt{\mu})$. 
First, we should observe that the contribution at  $T'\to\infty$ in Eq.~\eqref{eq:s1} 
vanishes. However, the contribution of $\mu\epsilon$ diverges in the 
limit $\epsilon\to0$. We anticipate that this divergence is regularised by 
the contribution of ${\mathcal C}^\epsilon$ (see Fig.~\ref{fig:app}). 
Precisely, one has 
\begin{equation}
	\label{eq:id}
	\Gamma(\omega,x)\approx\Gamma(\omega)-\frac{x^{\omega}}{\omega}
	+\frac{x^{\omega+1}}{1+\omega},\quad
	\mathrm{for}\,x\to0.
\end{equation}
Note that the number of singular terms depends on $\omega$. 
The first correction to the saddle point result can be easily derived, 
yielding  
\begin{multline}
\label{eq:s2}
{\mathcal I}_{{\mathcal C}^+}^{(1)}=
\frac{e^{-\frac{1}{2}(3-\gamma_m)\mu}2^m}{2\pi i(-1)^\omega \mu^2}
\Big\{
\frac{2^{m-1}3\gamma_m}{(8+2^m 5)}
\Gamma\Big(2-\omega,\frac{2^{1-m}u}{\gamma_m}\Big)+\\
\frac{\left(2^m 5+4\right) \left(3\ 2^m \gamma_m+2^m 5+8\right)}{2^4(8+2^m 5)}
\Gamma\Big(3-\omega,\frac{2^{1-m}u}{\gamma_m}\Big)
\Big)
\left.\Big\}\right|_{\mu\epsilon}^{\mu T'}\\\times
\left(\frac{2^{1-m}\mu}{\gamma_m}\right)^{\omega}
+{\mathcal O}(\mu^{\omega-3}). 
\end{multline}
%
%
We now observe that in both~\eqref{eq:s1} and~\eqref{eq:s2} in the limit 
$T'\to\infty$, we have $\Gamma(\omega,T')\to0$. For $\epsilon\to0$ similar divergences 
as for~\eqref{eq:s1} arise, which are removed by the integral on 
${\mathcal C}^\epsilon$. 

Before discussing the integral on ${\mathcal C}^\epsilon$, we focus 
on ${\mathcal I}_{{\mathcal C}^-}$ (see~\eqref{eq:int-ex}). 
The calculation is similar, the 
only difference is the phase factor due to the presence of the 
branch cut. Precisely, one has 
\begin{equation}
\label{eq:id1}
{\mathcal I}_{{\mathcal C}^-}=-{\mathcal I}_{{\mathcal C}^+}e^{-2\pi\omega i}. 
\end{equation}
From that~\eqref{eq:def-1}, for even $m$ one obtains that ${\mathcal 
I}_{{\mathcal C}^+}={\mathcal I}_{{\mathcal C}^-}$ (whereas for odd 
$m$ the two integrals cancel out). 

Finally, we briefly discuss the integral on the inner circle ${\mathcal C}^\epsilon$ 
around the branch cut. One has 
$u=\epsilon e^{i\theta}$. Therefore the integral to evaluate is 
\begin{multline}
\label{eq:epsi}
{\mathcal I}_{{\mathcal C}^\epsilon}=\frac{\epsilon^{1-\omega}}{4\pi(-1)^
\omega}\int^{2\pi}_0d\theta 
\frac{e^{-\frac{1}{2}(3-\sqrt{5+2^{3-m}e^{-\epsilon e^{i\theta}}})\mu}}{
	e^{i(\omega-1)\theta}}\\
	\times \Big[1+\frac{3}{\sqrt{5+2^{3-m}e^{-\epsilon e^{i\theta}}}}\Big]. 
\end{multline}
Since we are interested in the limit $\epsilon\to 0$, we can expand 
the integrand. After performing the integral over $\theta$, we obtain 
that at the leading order in $\epsilon$ one has  
\begin{equation}
\label{eq:int-eps}
{\mathcal I}_{{\mathcal C}^\epsilon}=
\frac{\epsilon^{1-\omega}e^{-2i\pi\omega}}
	{2\pi(\omega-1)} 
	\Big(1+\frac{3}{\gamma_m}
	\Big)e^{-\frac{1}{2}(3-\gamma_m)\mu}\sin(\pi\omega). 
\end{equation}
At the leading order  the contribution of~\eqref{eq:int-eps} 
cancels the most divergent term in ${\mathcal I}^{(0)}_{{\mathcal C}^+}
+{\mathcal I}^{(0)}_{{\mathcal C}^-}$ (see~\eqref{eq:s1}). We 
checked that higher order terms cancel higher order divergences 
in~\eqref{eq:s1} and~\eqref{eq:s2}. 

It is now straightforward to derive the result for the integral 
$I_m$ in~\eqref{eq:integral}. We focus on the leading order in 
$\mu$. At the leading order $\omega=(m+1)/2$,  
from~\eqref{eq:s1},~\eqref{eq:id1}, and the expression 
for $Q_0$ (cf.~\eqref{eq:q0}) one obtains that 
\begin{multline}
\label{eq:final}
I_m=
\frac{1}{2}e^{-\frac{1}{2}(3-\gamma_m)\mu}\Big(1+\frac{3}{\gamma_m}\Big)\Big(\frac{2}{\pi}\Big)^{\frac{1-m}{2}}
\Big(\frac{2^{1-m}\mu}{\gamma_m}\Big)^{\frac{m-1}{2}}.  
\end{multline}
Here we also used that the factor $\Gamma((m+1)/2)$ in~\eqref{eq:laplace} 
cancels out with the $\Gamma((1-m)/2)$ obtained from~\eqref{eq:s1} 
in the limit $\epsilon\to0$ (see~\eqref{eq:id}) using Euler's reflection 
formula 
\begin{equation}
\Gamma(x)\Gamma(1-x)=\frac{\pi}{\sin(\pi x)},\quad x\notin\mathbb{Z}. 
\end{equation}

Finally, we stress that the result \eqref{eq:final} coincides, at least to leading order, with the result in the main text \eqref{eq:4.10.v1}
obtained by taking the derivative of $g(\s)$ for odd $m$.




\begin{thebibliography}{99}


\bibitem{Amico2007}
L.~Amico, R.~Fazio, A.~Osterloh and V.~Vedral,
\newblock {\it Entanglement in many-body systems},
\newblock \href{http://dx.doi.org/10.1103/RevModPhys.80.517}{Rev. Mod. Phys. \textbf{80}, 517 (2008)}.

\bibitem{Calabrese2009R}
P.~Calabrese, J.~Cardy and B.~Doyon, 
\newblock {\it Entanglement entropy in extended quantum systems},
\newblock \href{http://dx.doi.org/10.1088/1751-8121/42/50/500301}{J. Phys. A  \textbf{42}, 500301 (2009)}.

\bibitem{Eisert2010}
J.~Eisert, M.~Cramer and M.~B.~Plenio,
\newblock {\it Area laws for the entanglement entropy}
\newblock \href{https://doi.org/10.1103/RevModPhys.82.277}{Rev. Mod. Phys. \textbf{82}, 277 (2010)}.

\bibitem{Laflorencie2015}
N.~Laflorencie,
\newblock {\it Quantum entanglement in condensed matter systems},
\newblock \href{http://dx.doi.org/10.1016/j.physrep.2016.06.008}{Phys. Rep.  \textbf{646}, 1 (2016)}.

\bibitem{Vidal2003}
G.~Vidal, J.I.~Latorre, E.~Rico and A.~Kitaev, 
\newblock {\it Entanglement in quantum critical phenomena}, 
\newblock \href{http://dx.doi.org/10.1103/PhysRevLett.90.227902}{Phys. Rev. Lett. \textbf{90}, 227902 (2003)}.

\bibitem{Latorre2004}
J. I.~Latorre, E.~Rico and G.~Vidal,
\newblock {\it Ground state entanglement in quantum spin chains},
\href{https://arxiv.org/abs/quant-ph/0304098}{Quant. Inf. Comp. \textbf{4}, 048 (2004)}.

\bibitem{Calabrese2004}
P.~Calabrese and J.~Cardy,
\newblock {\it Entanglement entropy and quantum field theory}, 
\newblock \href{http://dx.doi.org/10.1088/1742-5468/2004/06/P06002}{J. Stat. Mech. \textbf{2004}, P06002 (2004)}.

\bibitem{Calabrese2009}
P. Calabrese and J. Cardy, 
\newblock {\it Entanglement entropy and conformal field theory}, 
\newblock \href{http://dx.doi.org/10.1088/1751-8113/42/50/504005}{J. Phys. A \textbf{42}, 504005 (2009)}.



\bibitem{Li2008}
H.~Li and F.~D.~M. Haldane,
\newblock {\it Entanglement spectrum as a generalization of entanglement entropy: Identification of Topological Order in Non-Abelian Fractional Quantum Hall Effect States},
\newblock \href{http://dx.doi.org/10.1103/PhysRevLett.101.010504}{Phys. Rev. Lett. \textbf{101}, 010504 (2008)}.

\bibitem{Lefevre2008}
P.~Calabrese and A.~Lefevre,
\newblock {\it Entanglement spectrum in one-dimensional systems},
\newblock \href{http://dx.doi.org/10.1103/PhysRevA.78.032329}{Phys. Rev. A \textbf{78}, 032329 (2008)}.

\bibitem{Pollmann2010}
F.~Pollmann and J.~E. Moore,
\newblock {\it Entanglement spectra of critical and near-critical systems in one dimension},
\newblock \href{http://dx.doi.org/10.1088/1367-2630/12/2/025006}{New J. Phys. \textbf{12}, 025006 (2010)}.


\bibitem{Refael2004}
G.~Refael and J.~E.~Moore, 
\newblock {\it Entanglement Entropy of Random Quantum Critical Points in One Dimension},
\newblock \href{http://dx.doi.org/10.1103/PhysRevLett.93.260602}{Phys. Rev. Lett. {\bf 93}, 260602 (2004)}.

\bibitem{Laflorencie2005}
N.~Laflorencie, 
\newblock {\it Scaling of entanglement entropy in the random singlet phase},
\newblock \href{http://dx.doi.org/10.1103/PhysRevB.72.140408}{Phys. Rev. B {\bf 72}, 140408(R) (2005)}.

\bibitem{Dechiara2006}
G.~De Chiara, S.~Montangero, P.~Calabrese, and R.~Fazio,
 {\it Entanglement entropy dynamics of Heisenberg chains},
 \href{http://dx.doi.org/10.1088/1742-5468/2006/03/P03001}{J. Stat. Mech.  P03001 (2006)}.

\bibitem{Refael2007}
G.~Refael and J.~E.~Moore, 
\newblock {\it Entanglement entropy of the random s=1 Heisenberg chain},
\newblock \href{http://dx.doi.org/10.1103/PhysRevB.76.024419}{Phys. Rev. B {\bf 76}, 024419 (2007)}.

\bibitem{Binosi2007}
D. Binosi, G. De Chiara, S. Montangero, and A. Recati,
 {\it Increasing entanglement through engineered disorder in the random Ising chain},
 \href{http://dx.doi.org/10.1103/PhysRevB.76.140405}{Phys. Rev. B {\bf 76}, 140405 (2007)}.

\bibitem{Hoyos2007}
J.~A.~Hoyos, A.~P.~Vieira, N.~Laflorencie and E.~Miranda,
\newblock {\it Correlation amplitude and entanglement entropy in random spin chains},
\newblock \href{http://dx.doi.org/10.1103/PhysRevB.76.174425}{Phys. Rev. B {\bf 76}, 174425 (2007)}.


\bibitem{Bonesteel2007}
N.~E.~Bonesteel and K.~Yang, 
\newblock {\it Infinite-Randomness Fixed Points for Chains of Non-Abelian Quasiparticles},
\newblock \href{http://dx.doi.org/10.1103/PhysRevLett.99.140405}{Phys. Rev. Lett. {\bf 99}, 140405 (2007)}.

\bibitem{Refael2009}
G.~Refael and J.~E.~Moore,
\newblock {\it Criticality and entanglement in random quantum systems},
\newblock \href{http://dx.doi.org/10.1088/1751-8113/42/50/504010}{J. Phys. A  {\bf 42}, 504010 (2009)}.

\bibitem{Getelina2016}
J. C. Getelina, F. C. Alcaraz, and J. A. Hoyos, 
 {\it Entanglement properties of correlated random spin chains and similarities with conformally invariant systems},
 \href{http://dx.doi.org/10.1103/PhysRevB.93.045136}{Phys. Rev. B {\bf 93}, 045136 (2016)}.

\bibitem{Igloi2008}
F. Igloi and Y.-C. Lin,
 {\it Finite-size scaling of the entanglement entropy of the quantum Ising chain with homogeneous, periodically modulated and random couplings},
 \href{http://dx.doi.org/10.1088/1742-5468/2008/06/P06004}{J. Stat. Mech. P06004 (2008)}.

\bibitem{Fagotti2011}
M.~Fagotti, P.~Calabrese and J.~E.~Moore, 
\newblock {\it Entanglement spectrum of random-singlet quantum critical points},
\newblock \href{http://dx.doi.org/10.1103/PhysRevB.83.045110}{Phys. Rev. B {\bf 83},  045110 (2011)}.

\bibitem{Ramirez2014}
G.~Ramirez, J.~Rodriguez-Laguna, and G.~Sierra, 
{\it Entanglement in low-energy states of the random-hopping model},
\href{http://dx.doi.org/10.1088/1742-5468/2014/07/P07003}{J. Stat. Mech.  P07003 (2014)}.

\bibitem{trc-19}
X. Turkeshi, P. Ruggiero, and P. Calabrese, {\it Negativity Spectrum in the Random Singlet Phase},
\href{http://dx.doi.org/10.1103/PhysRevB.101.064207}{Phys. Rev. B {\bf 101}, 064207 (2020)}.

\bibitem{ruggiero-randneg}
P. Ruggiero, V. Alba, and P. Calabrese, {\it Entanglement negativity in random spin chains},
\newblock \href{https://journals.aps.org/prb/abstract/10.1103/PhysRevB.94.035152}{Phys. Rev. B {\bf 94}, 035152 (2016)}.
 
\bibitem{Raul2006}
R.~Santachiara,
\newblock {\it Increasing of entanglement entropy from pure to random quantum critical chains},
\newblock \href{http://dx.doi.org/10.1088/1742-5468/2006/06/L06002}{J. Stat. Mech.  \textbf{2006}, L06002 (2006)}. 

\bibitem{Fidkowski2008}
L.~Fidkowski, G.~Refael, N.~E.~Bonesteel and J.~E.~Moore,
\newblock {\it c-theorem violation for effective central charge of infinite-randomness fixed points}
\newblock \href{http://dx.doi.org/10.1103/PhysRevB.78.224204}{Phys. Rev. B {\bf 78}, 224204 (2008)}.

\bibitem{act-17}
V. Alba, P. Calabrese, and E. Tonni, 
{\it Entanglement spectrum degeneracy and Cardy formula in 1+1 dimensional conformal field theories},
 \href{http://dx.doi.org/10.1088/1751-8121/aa9365}{J. Phys. A {\bf 51}, 024001 (2018)}.

\bibitem{cc-05}
P. Calabrese and J. Cardy, 
{\it Evolution of Entanglement Entropy in One-Dimensional Systems},
 \href{http://dx.doi.org/10.1088/1742-5468/2005/04/P04010}{J. Stat. Mech. (2005) P04010}.


\bibitem{Islam2015}
R.~Islam, R.~Ma, P.~M. Preiss, M.~E. Tai, A.~Lukin, M.~Rispoli and M.~Greiner,
\newblock {\it Measuring entanglement entropy in a quantum many-body system},
\newblock \href{http://dx.doi.org/10.1038/nature15750}{Nature \textbf{528}, 77 (2015)}.

\bibitem{Kaufman2016}
A. M.~Kaufman, M.E.~Tai, A.~Lukin, M.~Rispoli, R.~Schittko, P. M.~Preiss and M.~Greiner,
\newblock {\it Quantum thermalization through entanglement in an isolated  many-body system},
\newblock \href{http://dx.doi.org/10.1126/science.aaf6725}{Science \textbf{353}, 764 (2016)}. 
 
\bibitem{Elben2018}
A.~Elben, B.~Vermersch, M.~Dalmonte, J.I.~Cirac and P.~Zoller, 
\newblock {\it R\'enyi Entropies from Random Quenches in Atomic Hubbard and Spin Models}, 
\newblock \href{http://dx.doi.org/10.1103/PhysRevLett.120.050406}{Phys. Rev. Lett. \textbf{120}, 050406 (2018)}.

\bibitem{Brydges2019}
T.~Brydges, A.~Elben, P.~Jurcevic, B.~Vermersch, C.~Maier, B.P.~Lanyon, P.~Zoller, R.~Blatt and C.F.~Roos,
\newblock {\it Probing R\'enyi entanglement entropy via randomized measurements},
\newblock \href{https://doi.org/10.1126/science.aau4963}{Science {\bf 364}, 6437 (2019)}.

\bibitem{Lukin2019}
A.~Lukin, M.~Rispoli, R.~Schittko, M.E.~Tai, A.M.~Kaufman, S.~Choi, V.~Khemani, J.~Leonard and M.Z.~Greiner,
\newblock {\it Probing entanglement in a many-body localized system},
\newblock \href{http://dx.doi.org/10.1126/science.aau0818}{Science \textbf{364}, 6437 (2019)}.

\bibitem{Lauchli2013}
A.~M. L\"auchli,
\newblock {\it Operator content of real-space entanglement spectra at conformal critical points},
\newblock \href{http://arxiv.org/abs/1303.0741}{arXiv:1303.0741 (2013)}.

\bibitem{Laflorencie2014}
N.~Laflorencie and S.~Rachel, 
\newblock {\it Spin-resolved entanglement spectroscopy of critical spin chains and Luttinger liquids},
\newblock \href{http://dx.doi.org/10.1088/1742-5468/2014/11/P11013}{J. Stat. Mech. P11013 (2014)}.

\bibitem{Xavier2018}
J.C.~Xavier, F.C.~Alcaraz and G.~Sierra, 
\newblock {\it Equipartition of the entanglement entropy}, 
\newblock \href{https://doi.org/10.1103/PhysRevB.98.041106}{Phys. Rev. B \textbf{98}, 041106 (2018)}.

\bibitem{Murciano2019}
S.~Murciano, G.~Di~Giulio and P.~Calabrese,
\newblock {\it Symmetry resolved entanglement in gapped integrable systems: a corner transfer matrix approach},
\newblock \href{https://doi.org/10.21468/SciPostPhys.8.3.046}{SciPost Phys. {\bf 8}, 046 (2020)}.

\bibitem{Goldstein2018}
M.~Goldstein and E.~Sela, 
\newblock {\it Symmetry-Resolved Entanglement in Many-Body Systems}, 
\newblock \href{https://doi.org/10.1103/PhysRevLett.120.200602}{Phys. Rev. Lett. \textbf{120}, 200602 (2018)}.

\bibitem{Goldstein2018B}
M.~Goldstein and E.~Sela, 
\newblock {\it Imbalance Entanglement: Symmetry Decomposition of Negativity}, 
\newblock \href{https://doi.org/10.1103/PhysRevA.98.032302}{Phys. Rev. A \textbf{98}, 032302 (2018)}.



\bibitem{Feld2019}
N~Feldman and M.~Goldstein,
\newblock {\it Dynamics of Charge-Resolved Entanglement after a Local Quench},
\newblock \href{https://doi.org/10.1103/PhysRevB.100.235146}{Phys. Rev. \textbf{B} 100, 235146 (2019)}.

\bibitem{Calabrese2020}
P.~Calabrese, M.~Collura, G.~Di~Giulio and S.~Murciano,
\newblock {\it Full counting statistics in the gapped XXZ spin chain},
\href{https://doi.org/10.1209/0295-5075/129/60007}{EPL {\bf 129}, 60007 (2020)}.

\bibitem{Bonsignori2019}
R.~Bonsignori, P.~Ruggiero and P.~Calabrese,
\newblock {\it Symmetry resolved entanglement in free fermionic systems},
\newblock \href{https://doi.org/10.1088/1751-8121/ab4b77}{J. Phys. A \textbf{52}, 475302 (2019)}.


\bibitem{Fraenkel2019}
S.~Fraenkel and M.~Goldstein,
\newblock {\it Symmetry resolved entanglement: Exact results in 1D and beyond},
\newblock \href{https://doi.org/10.1088/1742-5468/ab7753}{J. Stat. Mech. (2020) 033106}.


\bibitem{crc-20}
L. Capizzi, P. Ruggiero, and P. Calabrese, {\it Symmetry resolved entanglement entropy of excited states in a CFT}, 
\href{https://arxiv.org/abs/2003.04670}{arXiv:2003.04670 (2020)}.

\bibitem{sara2D}
S. Murciano, P. Ruggiero, and P. Calabrese, {\it Symmetry resolved entanglement in two-dimensional systems via dimensional reduction},
\href{https://arxiv.org/abs/2003.11453}{arXiv:2003.11453 (2020)}.

\bibitem{neg2}
 H. Shapourian, P. Ruggiero, S. Ryu, and P. Calabrese, {\it Twisted and untwisted negativity spectrum of free fermions}, 
 \href{http://dx.doi.org/10.21468/SciPostPhys.7.3.037}{SciPost Phys. {\bf 7}, 037 (2019)}.
 
\bibitem{clss-19}
E. Cornfeld, L. A. Landau, K. Shtengel, and E. Sela, 
{\it Entanglement spectroscopy of non-Abelian anyons: Reading off quantum dimensions of individual anyons},
\href{http://dx.doi.org/10.1103/PhysRevB.99.115429}{Phys. Rev. B {\bf 99}, 115429 (2019)}.

\bibitem{cms-13}
P. Caputa, G. Mandal, and R. Sinha, 
{\it Dynamical entanglement entropy with angular momentum and U(1) charge},
\href{https://dx.doi.org/10.1007/JHEP11(2013)052}{JHEP 11 (2013) 052}.

\bibitem{d-16}
J. S. Dowker,  
{\it Conformal weights of charged R\'enyi entropy twist operators for free scalar fields in arbitrary dimensions},
\href{https://doi.org/10.1088/1751-8113/49/14/145401}{J. Phys. A {\bf 49}, 145401 (2016)};\\
J. S. Dowker,  
{\it Charged R\'enyi entropies for free scalar fields},
\href{https://doi.org/10.1088/1751-8121/aa6178}{J. Phys. A {\bf 50}, 165401 (2017)}.

\bibitem{matsuura}
A. Belin, L.-Y. Hung, A. Maloney, S. Matsuura, R. C. Myers, and T. Sierens,
{\it Holographic charged R\'enyi entropies}, 
\href{https://doi.org/10.1007/JHEP12(2013)059}{JHEP {\bf 12} (2013) 059}.


\bibitem{SREE}
P. Caputa, M. Nozaki, and T. Numasawa, 
{\it Charged Entanglement Entropy of Local Operators}, 
\href{https://dx.doi.org/10.1103/PhysRevD.93.105032}{Phys. Rev. D {\bf 93}, 105032 (2016)}.





\bibitem{Ma1979}       
S.-K.~Ma, C.~Dasgupta and C.-k.~Hu,
\newblock {\it Random Antiferromagnetic Chain},
\newblock \href{http://dx.doi.org/10.1103/PhysRevLett.43.1434}{Phys. Rev. Lett. \textbf{43}, 1434 (1979)}.
       
\bibitem{Ma1980}
C.~Dasgupta and S.-K.~Ma,
\newblock {\it Low-temperature properties of the random Heisenberg antiferromagnetic chain},
\newblock \href{http://dx.doi.org/10.1103/PhysRevB.22.1305}{Phys. Rev. B \textbf{22}, 1305 (1980)}.

\bibitem{Fisher1994}
D.~Fisher,
\newblock {\it Random antiferromagnetic quantum spin chains},
\newblock \href{http://dx.doi.org/10.1103/PhysRevB.50.37990}{Phys. Rev. B \textbf{50}, 3799 (1994)}.
 
\bibitem{Igloi05}
F. Igl\'oi and C. Monthus, 
\newblock {\it Strong disorder RG approach of random systems},
\newblock \href{http://dx.doi.org/10.1016/j.physrep.2005.02.006}{Phys. Rep. \textbf{412}, 277 (2005)}.

\bibitem{Monthus2018}
F.~Igl\'oi and C.~Monthus
\newblock {\it Strong Disorder RG approach - a short review of recent developments},
\newblock \href{http://dx.doi.org/10.1140/epjb/e2018-90434-8}{Eur. Phys. J. B \textbf{91}, 290 (2018)}.

\bibitem{Doty1992}
C.~A.~Doty and D.~S.~Fisher
\newblock {\it Effects of quenched disorder on spin-1/2 quantum XXZ chains},
\newblock \href{https://doi.org/10.1103/PhysRevB.45.2167}{Phys. Rev. B \textbf{45}, 2167 (1992)}.


\bibitem{nc-10}
M.~A. Nielsen and I.~L. Chuang, \textit{{Quantum computation and quantum  information}}.
 \href{http://dx.doi.org/10.1017/CBO9780511976667}{Cambridge University Press, Cambridge, UK, 10th anniversary~ed. (2010)}.


\bibitem{klick2009}
I. Klich and L. Levitov, {\it Quantum Noise as an Entanglement Meter},
\newblock \href{https://journals.aps.org/prl/abstract/10.1103/PhysRevLett.102.100502}{Phys. Rev. Lett. {\bf 102}, 100502 (2009)}.

\bibitem{song2010}
H. F. Song, S. Rachel, and K. Le Hur, {\it General relation between entanglement and fluctuations in one dimension},
\newblock \href{https://journals.aps.org/prb/abstract/10.1103/PhysRevB.82.012405}{Phys. Rev. B {\bf 82}, 012405 (2010)}.

\bibitem{song2011}
H. F. Song, C. Flindt, S. Rachel, I. Klich, and K. Le Hur, {\it Entanglement entropy from charge statistics: Exact relations for noninteracting many-body systems},
\newblock \href{https://journals.aps.org/prb/abstract/10.1103/PhysRevB.83.161408}{Phys. Rev. B {\bf 83}, 161408(R) (2011)}.

\bibitem{song2012}
H. F. Song, S.~Rachel, C. Flindt, I. Klich, N. Laflorencie, and K. Le Hur, {\it Bipartite fluctuations as a probe of many-body entanglement},
\newblock \href{https://journals.aps.org/prb/abstract/10.1103/PhysRevB.85.035409}{Phys. Rev. B {\bf 85}, 035409 (2012)}.

\bibitem{rachel2012}
S. Rachel, N. Laflorencie, H. F. Song, and K. Le Hur, {|it Detecting Quantum Critical Points Using Bipartite Fluctuations},
\newblock \href{https://journals.aps.org/prl/abstract/10.1103/PhysRevLett.108.116401}{Phys. Rev. Lett. {\bf 108}, 116401 (2012)}.


\bibitem{cmv-12}
P. Calabrese, M. Mintchev and E. Vicari, {\it Exact relations between particle fluctuations and entanglement in Fermi gases},
\href{http://dx.doi.org/10.1209/0295-5075/98/20003}{EPL {\bf 98}, 20003 (2012).}
 
\bibitem{si-13}
R. Susstrunk and D. A. Ivanov,
{\it Free fermions on a line: Asymptotics of the entanglement entropy and entanglement spectrum from full counting statistics},
\href{https://doi.org/10.1209/0295-5075/100/60009}{EPL {\bf 100}, 60009 (2012).}

\bibitem{clm-15}
P. Calabrese, P. Le Doussal, and S. N. Majumdar, {\it Random matrices and entanglement entropy of trapped Fermi gases},
\href{https://doi.org/10.1103/PhysRevA.91.012303}{Phys. Rev. A {\bf 91}, 012303 (2015).}

\bibitem{petrescu2014}
A. Petrescu, H. F. Song, S. Rachel, Z. Ristivojevic, C. Flindt, N. Laflorencie, I. Klich, N. Regnault and K. Le Hur,
{\it Fluctuations and entanglement spectrum in quantum Hall states},
\newblock \href{https://doi.org/10.1088/1742-5468/2014/10/P10005}{J. Stat. Mech. (2014) P10005}.


\bibitem{wv-03}
H. M. Wiseman and J. A. Vaccaro, 
{\it Entanglement of Indistinguishable Particles Shared between Two Parties}, 
\href{https://dx.doi.org/10.1103/PhysRevLett.91.097902}{ Phys. Rev. Lett. {\bf 91}, 097902 (2003)}.


\bibitem{SREE2d}
H. Barghathi, C. M. Herdman, and A. Del Maestro, 
{\it R\'enyi generalization of the operational entanglement entropy}, 
\href{http://dx.doi.org/10.1103/PhysRevLett.121.150501}{Phys. Rev. Lett. {\bf 121}, 150501 (2018)}.

\bibitem{delmaestro2}
H. Barghathi, E. Casiano-Diaz, and A. Del Maestro,
{\it Operationally accessible entanglement of one-dimensional spinless fermions}, 
\href{https://journals.aps.org/pra/abstract/10.1103/PhysRevA.100.022324}{Phys. Rev. A {\bf 100}, 022324 (2019)}.

\bibitem{kusf-20}
M. Kiefer-Emmanouilidis, R. Unanyan, J. Sirker, and M. Fleischhauer, 
{\it Bounds on the entanglement entropy by the number entropy in non-interacting fermionic systems},
\href{https://arxiv.org/pdf/2003.03112.pdf}{arXiv:2003.03112}.

\bibitem{kusf-20b}
M. Kiefer-Emmanouilidis, R. Unanyan, J. Sirker, and M. Fleischhauer,
{\it Evidence for unbounded growth of the number entropy in many-body localized phases},
\href{https://arxiv.org/pdf/2003.04849.pdf}{arXiv:2003.04849}.



\bibitem{Lin2007}
Y.-C.~Lin, F.~Igloi and H.~Rieger, 
\newblock {\it Entanglement Entropy at Infinite-Randomness Fixed Points in Higher Dimensions},
\newblock \href{http://dx.doi.org/10.1103/PhysRevLett.99.147202}{Phys. Rev. Lett. {\bf 99}, 147202 (2007)}.

\bibitem{Yu2008}
R. Yu, H. Saleur, and S. Haas,  
 {\it Entanglement entropy in the two-dimensional random transverse field Ising model},
 \href{http://dx.doi.org/10.1103/PhysRevB.77.140402}{Phys. Rev. B {\bf 77}, 140402 (2008)}.
 
\bibitem{Kovaks2009}
I. A. Kovacs and F. Igloi, 
 {\it Critical behavior and entanglement of the random transverse-field Ising model between one and two dimensions},
 \href{http://dx.doi.org/10.1103/PhysRevB.80.214416}{Phys. Rev. B {\bf 80}, 214416 (2009)}.

\bibitem{Kovaks2012}
I. A. Kovacs and F. Igloi, 
 {\it Universal logarithmic terms in the entanglement entropy of 2d, 3d and 4d random transverse-field Ising models},
 \href{http://dx.doi.org/10.1209/0295-5075/97/67009}{EPL {\bf 97}, 67009 (2012)}.

\bibitem{vjs-13}
R. Vasseur, J. L. Jacobsen, and H. Saleur, {\it Universal entanglement crossover of coupled quantum wires}, 
\href{http://dx.doi.org/10.1103/PhysRevLett.112.106601}{Phys. Rev. Lett. {\bf 112}, 106601 (2014)}.

\bibitem{Laguna2016}
J. Rodriguez-Laguna, S. N. Santalla, G. Ramirez, and G. Sierra, 
 {\it Entanglement in correlated random spin chains, RNA folding and kinetic roughening},
 \href{http://dx.doi.org/10.1088/1367-2630/18/7/073025}{New J. Phys. \textbf{18}, 073025 (2016)}.
 
 \bibitem{vrhs-17}
R. Vasseur, A. Roshani, S. Haas, and H. Saleur, {\it Healing of Defects in Random Antiferromagnetic Spin Chains},
 \href{http://dx.doi.org/10.1209/0295-5075/119/50004}{EPL {\bf 119}, 50004 (2017)}.



\bibitem{asr-18}
V. Alba, S. N. Santalla, P. Ruggiero, J. Rodriguez-Laguna, P. Calabrese, and G. Sierra, {\it Unusual area-law violation in random inhomogeneous systems}, 
\href{http://dx.doi.org/10.1088/1742-5468/ab02df}{J. Stat. Mech. (2019) 023105}.



\bibitem{pcp-19}
S. Pappalardi, P. Calabrese, and G. Parisi, {\it Entanglement entropy of the long-range Dyson hierarchical model},
\href{https://doi.org/10.1088/1742-5468/ab2903}{J. Stat. Mech. (2019) 073102}. 

\bibitem{m-20}
C. Monthus, {\it Properties of the simplest inhomogeneous and homogeneous Tree-Tensor-States for Long-Ranged Quantum Spin Chains with or without disorder},
\href{https://arxiv.org/abs/2001.10731}{arXiv:2001.10731}.



\bibitem{bss-07}
M. Bortz, J. Sato, and M. Shiroishi M, {\it String correlation functions of the spin-1/2 Heisenberg XXZ chain},  
\href{https://doi.org/10.1088/1751-8113/40/16/001}{J. Phys. A {\bf 40}, 4253 (2007)}.

\bibitem{aem-08}
D. B. Abraham, F. H. L. Essler, and A. Maciolek, {\it Effective Forces Induced by a Fluctuating Interface: Exact Results}, 
\href{https://doi.org/10.1103/PhysRevLett.98.170602}{Phys. Rev. Lett. {\bf 98}, 170602 (2007)}.

\bibitem{zpp-07}
M. Znidaric, T. Prosen, and P. Prelovsek, {\it Many body localization in Heisenberg XXZ magnet in a random field},
\href{https://doi.org/10.1103/PhysRevB.77.064426}{Phys. Rev. B {\bf 77}, 064426 (2008)}.

 \bibitem{Bardarson2012}
J. H. Bardarson, F. Pollmann, and J. E. Moore,
 {\it Unbounded Growth of Entanglement in Models of Many-Body Localization}, 
 \href{http://dx.doi.org/10.1103/PhysRevLett.109.017202}{Phys. Rev. Lett. {\bf 109}, 017202 (2012)}.

\bibitem{Serbyn13}
M. Serbyn, Z. Papic, and D. A. Abanin, 
 {\it Universal Slow Growth of Entanglement in Interacting Strongly Disordered Systems},
 \href{http://dx.doi.org/10.1103/PhysRevLett.110.260601}{Phys. Rev. Lett. {\bf 110}, 260601 (2013)}.

\bibitem{Vosk2014}
R. Vosk and E. Altman, 
 {\it Dynamical Quantum Phase Transitions in Random Spin Chains},
 \href{http://dx.doi.org/10.1103/PhysRevLett.112.217204}{Phys. Rev. Lett. {\bf 112}, 217204 (2014)}.

\bibitem{Altman15}
 E. Altman and R. Vosk, 
 {\it Universal dynamics and renormalization in many body localized systems},
 \href{http://dx.doi.org/10.1146/annurev-conmatphys-031214-014701}{Ann. Review of Cond. Mat. Phys. {\bf 6}, 383 (2015)}.

\bibitem{Parameswaran2017}
S. A. Parameswaran, A. C. Potter and R. Vasseur, 
 {\it Eigenstate phase transitions and the emergence of universal dynamics in highly excited states},
 \href{http://dx.doi.org/10.1002/andp.201770051}{Annalen der Physik  {\bf 529}, 1600302 (2017)}.

\bibitem{Abanin18}
D. A. Abanin, E. Altman, I. Bloch and M. Serbyn,
 {\it Ergodicity, Entanglement and Many-Body Localization},
 \href{http://dx.doi.org/10.1103/RevModPhys.91.021001}{Rev. Mod. Phys. \textbf{91}, 021001 (2019)}

\bibitem{Pekker2014}
D. Pekker, G. Refael, E. Altman, E. Demler, and V. Oganesyan,
 {\it Hilbert-Glass Transition: New Universality of Temperature-Tuned Many-Body Dynamical Quantum Criticality},
 \href{http://dx.doi.org/10.1103/PhysRevX.4.011052}{Phys. Rev. X {\bf 4}, 011052 (2014)}.

\bibitem{Zhao2016}
Y. Zhao, F. Andraschko, and J. Sirker,  
 {\it Entanglement entropy of disordered quantum chains following a global quench},
 \href{http://dx.doi.org/10.1103/PhysRevB.93.205146}{Phys. Rev. B \textbf{93}, 205146 (2016)}.

\bibitem{exp-mbl}
 B. Chiaro et. al., {\it Growth and preservation of entanglement in a many-body localized system}, 
 \href{http://arxiv.org/abs/1910.06024}{arXiv:1910.06024}. 



\end{thebibliography}
\end{document}